\documentclass{SciPost}

\hypersetup{
    colorlinks,
    linkcolor={red!50!black},
    citecolor={blue!50!black},
    urlcolor={blue!80!black}
}

\usepackage[bitstream-charter]{mathdesign}
\DeclareSymbolFont{usualmathcal}{OMS}{cmsy}{m}{n}
\DeclareSymbolFontAlphabet{\mathcal}{usualmathcal}

\fancypagestyle{SPstyle}{
\fancyhf{}
\lhead{\colorbox{scipostblue}{\bf \color{white} ~SciPost Physics Codebases }}
\rhead{{\bf \color{scipostdeepblue} ~Submission }}

\fancyfoot[C]{\textbf{\thepage}}
}

\usepackage{placeins}
\usepackage{listings}
\definecolor{darkblue}{RGB}{10,40,120}
\usepackage{array,tabularx,seqsplit}
\newcolumntype{F}{>{\ttfamily}l}
\newcolumntype{T}{>{\raggedright\arraybackslash}X} 
\newcolumntype{M}{>{\raggedright\arraybackslash}X}

\newcommand{\tcode}[1]{\begingroup\ttfamily\small\seqsplit{#1}\endgroup}

\DeclareRobustCommand{\ttcode}[1]{\begingroup\normalfont\ttfamily\seqsplit{#1}\endgroup}

\newcommand{\bra}[1]{\langle#1|}
\newcommand{\ket}[1]{|#1\rangle}

\begin{document}

\pagestyle{SPstyle}

\begin{center}{\Large \textbf{\color{scipostdeepblue}{
ParaToric 1.0: Continuous-time quantum Monte Carlo \\ for the toric code in a parallel field\\
}}}\end{center}

\begin{center}\textbf{
Simon M. Linsel\textsuperscript{$\star$} and
Lode Pollet\textsuperscript{$\dagger$}
}\end{center}

\begin{center}
Department of Physics and Arnold Sommerfeld Center for Theoretical Physics (ASC), Ludwig-Maximilians-Universit\"at M\"unchen, Theresienstr. 37, D-80333, M\"unchen, Germany
\\
Munich Center for Quantum Science and Technology (MCQST), Schellingstr. 4, D-80799, M\"unchen, Germany
\\[\baselineskip]
$\star$ \href{mailto:simon.linsel@lmu.de}{\small simon.linsel@lmu.de}\,,\quad
$\dagger$ \href{mailto:lode.pollet@lmu.de}{\small lode.pollet@lmu.de}
\end{center}

\section*{\color{scipostdeepblue}{Abstract}}
\textbf{\boldmath{%
We introduce ParaToric, a C++ package for simulating the toric code in a parallel field (i.e., $X$- and $Z$-fields) at finite temperature. We implement and extend the continuous-time quantum Monte Carlo algorithm of Wu, Deng, and Prokof'ev on the square, triangular, honeycomb, and cubic lattices with either periodic or smooth open boundaries. The package is expandable to arbitrary lattice geometries and custom observables diagonal in either the $X$- or $Z$-basis. ParaToric also supports snapshot extraction in both bases, making it ideal for generating training/benchmarking data for other methods, such as lattice gauge theories, cold atom or other quantum simulators, quantum spin liquids, artificial intelligence, and quantum error correction. The software provides bindings to C/C++ and Python, and is thus almost universally integrable into other software projects. 
}}

\vspace{\baselineskip}

\noindent\textcolor{white!90!black}{%
\fbox{\parbox{0.975\linewidth}{%
\textcolor{white!40!black}{\begin{tabular}{lr}%
  \begin{minipage}{0.6\textwidth}%
    {\small Copyright attribution to authors. \newline
    This work is a submission to SciPost Physics Codebases. \newline
    License information to appear upon publication. \newline
    Publication information to appear upon publication.}
  \end{minipage} & \begin{minipage}{0.4\textwidth}
    {\small Received Date \newline Accepted Date \newline Published Date}%
  \end{minipage}
\end{tabular}}
}}
}


\vspace{10pt}
\noindent\rule{\textwidth}{1pt}
\tableofcontents
\noindent\rule{\textwidth}{1pt}
\vspace{10pt}

\section{Introduction}
\label{sec:intro}

The toric code is one of the most fundamental and most-studied models in modern condensed matter physics. It was first written down by Kitaev \cite{Kitaev2003} and is the simplest example of a model hosting a topological ground state (a gapped $\mathbb{Z}_2$ quantum spin liquid) and anyonic excitations. The toric code is also the foundational model for error-correcting codes \cite{Dennis2002, Fowler2012} and has deep connections to the Ising gauge theory \cite{Kogut1979}. It has been studied using a wide range of analytical and numerical techniques, including dual/exact mappings to spin models\footnote{The resulting spin models are typically solved with the numerical techniques mentioned below.} \cite{Trebst2007, Tupitsyn2010, Dusuel2010, Schmidt2013, Reiss2019, Kott2024}, series expansions \cite{Vidal2009, Vidal2009_2, Schulz2012, Kamfor2014, Reiss2019, Kott2024}, exact diagonalization \cite{Vidal2009_2, Morampudi2014, Schuler2016}, density matrix renormalization group calculations on cylinders \cite{Jiang2012, Jiang2013}, projected entangled pair states \cite{Dusuel2011, Schulz2012, Vanderstraeten2017, Crone2020, Iqbal2021, Xu2024}, quantum Monte Carlo \cite{Wu2012, Linsel2024, Linsel2026}, and neural networks \cite{Zhang2017, Valenti2022, Hibat-Allah2023, Kufel2025}. The toric code and its variants have also been experimentally realized using photonic systems \cite{Pachos2009, Liu2019}, nuclear magnetic resonance systems \cite{Feng2013, Park2016}, superconducting qubits \cite{Zhong2016, Song2018, Satzinger2021, Niu2024}, neutral atoms \cite{Bluvstein2022}, and trapped ions \cite{Iqbal2024, Iqbal2025}. For a review of the toric code, see \cite{Resende2020}.

One of the most common extensions of the toric code are parallel fields, which, when strong enough, destroy the topological order of the ground state. This model is sign-problem-free, thus making quantum Monte Carlo (QMC) the method of choice for studying equilibrium physics. Wu, Deng, and Prokof'ev developed a continuous-time quantum Monte Carlo algorithm \cite{Wu2012}. ParaToric implements and extends this algorithm with new updates which enable ergodicity at large temperatures and at zero off-diagonal field, thus significantly improving the applicability of the algorithm.

ParaToric implements a wide range of lattices, boundary conditions, and observables. It is also possible to extend ParaToric with new interactions, observables, and lattices. We provide documented interfaces in C, C++, and Python as well as command-line interfaces, making the integration of ParaToric into other projects and programming languages straightforward. ParaToric will save simulation results to HDF5 files and snapshots to GraphML files (XML-based), with a focus on interoperability with other packages. ParaToric comes with an MIT license.

\section{The toric code in a parallel field}

\subsection{Hamiltonian}

ParaToric implements and extends the continuous-time QMC algorithm of Wu, Deng, and Prokof'ev \cite{Wu2012} to simulate the toric code in a parallel field (also called perturbed toric code or extended toric code)
\begin{equation} \label{eq:eTC}
\hat{\mathcal{H}} = - \mu \sum_v \hat{A}_v \; - J \sum_{p} \hat{B}_p \; - h \sum_{l} \hat{\sigma}_l^x \; - \lambda \sum_{l} \hat{\sigma}_l^z,
\end{equation}
where $J, \lambda > 0$ in the $\hat{\sigma}^x$-basis and $\mu, h > 0$ in the $\hat{\sigma}^z$-basis (otherwise the model has a sign-problem). $\hat{\sigma}_l^x$ and $\hat{\sigma}_l^z$ are Pauli matrices defined on the links of the underlying lattice. The star term $\hat{A}_v$ contains all links adjacent to lattice site $v$, the plaquette term $\hat{B}_p$ contains all links that belong to the same elementary plaquette $p$ of the underlying lattice. The temperature $T = 1/\beta$ is finite. For readers interested in extending the code, we note that it is relatively straightforward to add interactions that are diagonal in the chosen basis, such as (long-range) Ising interactions. Off-diagonal interactions require a more careful review and extension of the Monte Carlo updates to ensure ergodicity. However, diagonal interactions can also cause sampling problems, especially when they introduce frustration, e.g., when flipping the sign of either the star or the plaquette term \cite{Borla2025}.

\subsection{Lattice geometries}

\begin{figure}[t]
\centering
\includegraphics[width=0.75\textwidth]{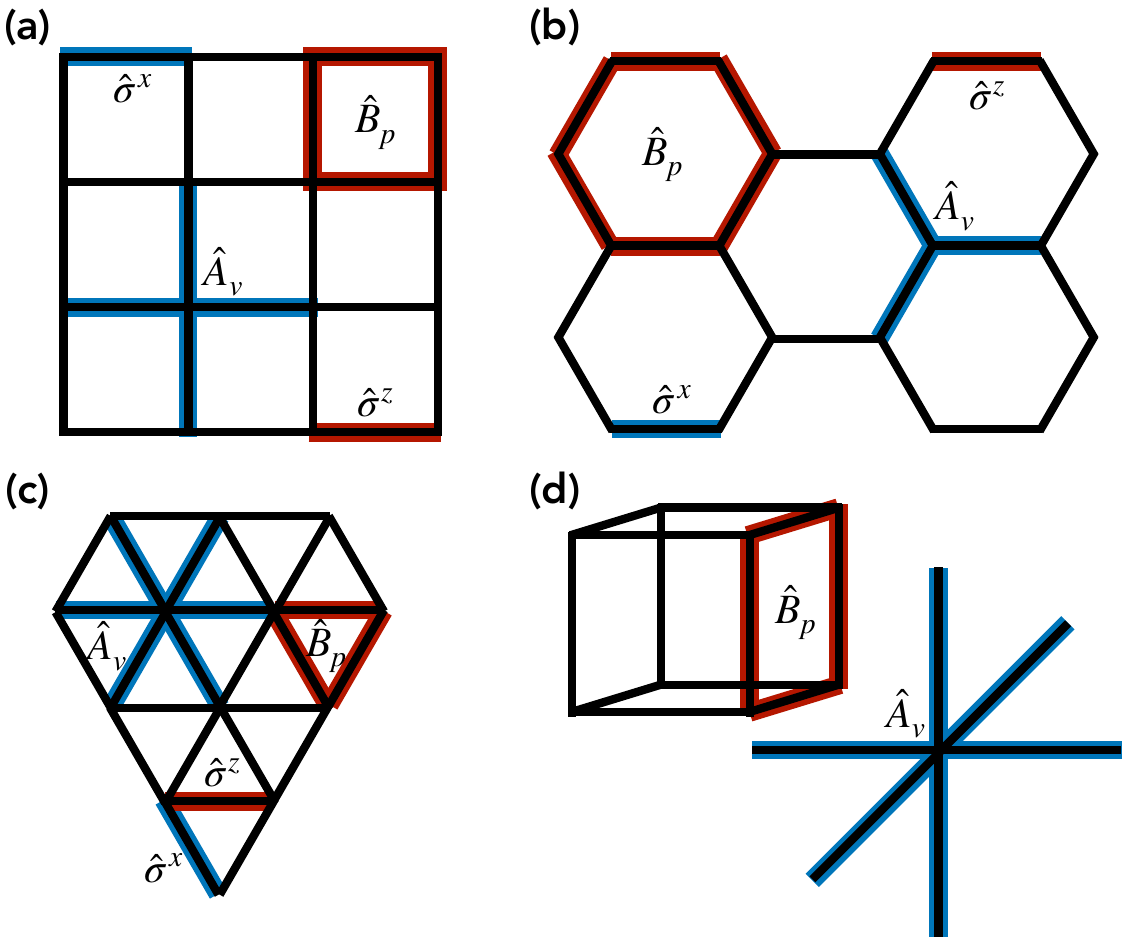}
\caption{\textbf{Implemented lattices}. We implement the extended toric code (\ref{eq:eTC}) on the square (a), honeycomb (b), triangular (c), and cubic lattices (d). For each lattice, we show the star ($\hat{A}_v$) and plaquette ($\hat{B}_p$) terms. The cubic lattice importantly features star interactions of six links and plaquette interactions of four links on the faces of cubes.}
\label{fig:TC}
\end{figure}

We implement the square, honeycomb, triangular, and cubic lattices, see Fig.~\ref{fig:TC}. On the cubic lattice, the plaquettes contain the four links of cube faces, \textit{not} the twelve links of the cube (that model has a different m-anyon structure). We implement periodic and smooth open (e.g., square lattice: four-qubit plaquettes at the boundaries, three-qubit stars at the boundaries except for the corners, two-qubit stars at the corners) boundaries, respectively. New lattices can be added in \ttcode{src/lattice/lattice.cpp}.

\subsection{Observables} \label{sec:observables}

Here we list all observables that ParaToric implements for the extended toric code. Custom observables can be added in \ttcode{src/mcmc/extended\_toric\_code\_qmc.hpp}. For each observable $\hat{O}$, we calculate the expectation value $\langle \hat{O} \rangle$ and the Binder ratio $U_O = \frac{\langle \hat{O}^4 \rangle}{\langle \hat{O}^2 \rangle ^2}$ with error bars obtained from bootstrapping (see below), respectively.

\paragraph{\ttcode{anyon\_count}} The number of e-anyons ($\hat{\sigma}^x$-basis) or m-anyons ($\hat{\sigma}^z$-basis) in the system.

\paragraph{\ttcode{anyon\_density}} $\hat{\sigma}^x$-basis: The number of e-anyons divided by the number of lattice sites. $\hat{\sigma}^z$-basis: The number of m-anyons divided by the number of plaquettes.

\paragraph{\ttcode{delta}} The difference between the star and plaquette expectation values: $\Delta = \langle \hat{A}_v \rangle - \langle \hat{B}_p \rangle$. 

\paragraph{\ttcode{energy}} The total energy $E=\langle \hat{\mathcal{H}} \rangle$.

\paragraph{\ttcode{energy\_h}} The electric field term $E_h=\langle - h \sum_{l} \hat{\sigma}_l^x \rangle$.

\paragraph{\ttcode{energy\_lmbda}} The gauge field term $E_{\lambda}=\langle - \lambda \sum_{l} \hat{\sigma}_l^z \rangle$. We write $\lambda$ as \ttcode{lmbda} because some programming languages feature \ttcode{lambda} as a keyword.

\paragraph{\ttcode{energy\_J}} The plaquette term $E_J=\langle - J \sum_{p} \hat{B}_p \rangle$.

\paragraph{\ttcode{energy\_mu}} The star term $E_{\mu}=\langle - \mu \sum_v \hat{A}_v \rangle$.

\paragraph{\ttcode{fredenhagen\_marcu}} The equal-time Fredenhagen-Marcu loop operator \cite{Fredenhagen1983, Fredenhagen1986, Fredenhagen1988}:
\begin{align}
    O_\mathrm{FM}^{x/z} = \lim_{L \to \infty} \frac{\langle \prod_{l \in \mathcal{C}^{x/z}_{1/2}} \hat{\sigma}^{x/z}_l \rangle}{\sqrt{|\langle \prod_{l \in \mathcal{C}^{x/z}} \hat{\sigma}^{x/z}_l \rangle|}},
\end{align}
$\mathcal{C}^{x/z}_{1/2}$ is half, $\mathcal{C}^{x/z}$ is a full Wilson loop in the $\hat{\sigma}^x$-basis ('t Hooft loop in the $\hat{\sigma}^z$-basis). The full Wilson loop contains the boundary links of a square with side length $L/2$ between $x/y=L/4$ and $x/y=3L/4$, where $L$ is the linear system size. It is automatically constructed for all supported lattices. When probing perimeter/area laws, the user should change $L$. We currently do not support off-diagonal loop operators, e.g., measuring products of $\hat{\sigma}^x$-operators in the $\hat{\sigma}^z$-basis.

\paragraph{\ttcode{largest\_cluster}} The largest connected cluster of neighboring bonds with $\hat{\sigma}^{x} = -1$ ($\hat{\sigma}^{z} = -1$) in the $\hat{\sigma}^x$-basis ($\hat{\sigma}^z$-basis). This observable is used to calculate the percolation strength, see \cite{Linsel2024}.

\paragraph{\ttcode{largest\_plaquette\_cluster}} The largest connected cluster of neighboring elementary plaquettes that share a bond with $\hat{\sigma}^{x} = -1$ ($\hat{\sigma}^{z} = -1$) in the $\hat{\sigma}^x$-basis ($\hat{\sigma}^z$-basis). This observable is used to calculate the plaquette percolation strength.

\paragraph{\ttcode{percolation\_probability}} Measures the bond percolation probability, i.e. if we can wind around the system while only traversing bonds with $\hat{\sigma}^{x} = -1$ ($\hat{\sigma}^{z} = -1$) in the $\hat{\sigma}^x$-basis ($\hat{\sigma}^z$-basis). Formally, it is the expectation value $\langle \hat{\Pi}^{x/z} \rangle$ of the projector
\begin{align}
    \hat{\Pi}^{x/z} = \sum_{ W(j) \neq 0} \ket{\{\hat{\sigma}^{x/z}\}_j} \bra{\{\hat{\sigma}^{x/z}\}_j},
\end{align}
over all possible configurations $\{\hat{\sigma}^{x/z}\}_j$ with non-zero winding number $W(j)$ of connected link clusters of neighboring $\hat{\sigma}^{x/z} = -1$. These clusters are called percolating clusters. For details, see \cite{Linsel2024, Duennweber2025, Linsel2026}. 

\paragraph{\ttcode{percolation\_strength}} If a snapshot does not have a percolating cluster, the percolation strength is 0. If a snapshot has a percolating cluster, the percolation strength is \ttcode{largest\_cluster} divided by the total number of links in the system. For details, see \cite{Linsel2024, Linsel2026}.

\paragraph{\ttcode{plaquette\_percolation\_probability}} Similar to the percolation probability of bonds. Two plaquettes are in the same cluster if they share a link $l$ with $\hat{\sigma}^x_l = -1$. For details, see \cite{Linsel2026}.  

\paragraph{\ttcode{plaquette\_percolation\_strength}} If a snapshot does not have a plaquette-percolating cluster, the plaquette percolation strength is 0. If a snapshot has a plaquette-percolating cluster, the plaquette percolation strength is \ttcode{largest\_plaquette\_cluster} divided by the total number of elementary plaquettes in the system. 

\paragraph{\ttcode{plaquette\_z}} The plaquette expectation value $\langle \hat{B}_p \rangle$.

\paragraph{\ttcode{sigma\_x}} The electric field expectation value $\langle \hat{\sigma}^x \rangle$.

\paragraph{\ttcode{sigma\_x\_static\_susceptibility}} The static susceptibility 
\begin{align}
    \chi_x^{\mathrm{stat}} = \int_0^\beta \langle \hat{\sigma}^x(0) \, \hat{\sigma}^x(\tau) \rangle_\mathrm{c} \; \mathrm{d}\tau,
\end{align}
where $\langle...\rangle_\mathrm{c}$ is the connected correlator, and the integral is over the imaginary time $\tau$. Importantly, $\chi_x^{\mathrm{stat}}$ can be calculated in both the $\hat{\sigma}^x$- and $\hat{\sigma}^z$-basis.

\paragraph{\ttcode{sigma\_x\_dynamical\_susceptibility}} The dynamical (fidelity) susceptibility 
\begin{align}
    \chi_x^{\mathrm{dyn}} = \int_0^{\beta/2} \tau \langle \hat{\sigma}^x(0) \, \hat{\sigma}^x(\tau) \rangle_\mathrm{c} \; \mathrm{d}\tau,
\end{align}
where $\langle...\rangle_\mathrm{c}$ is the connected correlator, and the integral is over the imaginary time $\tau$\footnote{Compared to the static susceptibility, the dynamical (fidelity) susceptibility contains an extra $\tau$ dependency in the integral.}. Importantly, $\chi_x^{\mathrm{dyn}}$ can be calculated in both the $\hat{\sigma}^x$- and $\hat{\sigma}^z$-basis. In the $\hat{\sigma}^z$-basis, we use Eq.~(9) of \cite{Wang2015}.

\paragraph{\ttcode{sigma\_z}} The gauge field expectation value $\langle \hat{\sigma}^z \rangle$.

\paragraph{\ttcode{sigma\_z\_static\_susceptibility}} The static susceptibility 
\begin{align}
    \chi_z^{\mathrm{stat}} = \int_0^\beta \langle \hat{\sigma}^z(0) \, \hat{\sigma}^z(\tau) \rangle_\mathrm{c}  \; \mathrm{d}\tau,
\end{align}
where $\langle...\rangle_\mathrm{c}$ is the connected correlator, and the integral is over the imaginary time $\tau$. Importantly, $\chi_z^{\mathrm{stat}}$ can be calculated in both the $\hat{\sigma}^x$- and $\hat{\sigma}^z$-basis.

\paragraph{\ttcode{sigma\_z\_dynamical\_susceptibility}} The dynamical (fidelity) susceptibility 
\begin{align}
    \chi_z^{\mathrm{dyn}} = \int_0^{\beta/2} \tau \langle \hat{\sigma}^z(0) \, \hat{\sigma}^z(\tau) \rangle_\mathrm{c} \; \mathrm{d}\tau,
\end{align}
where $\langle...\rangle_\mathrm{c}$ is the connected correlator, and the integral is over the imaginary time $\tau$. Importantly, $\chi_z^{\mathrm{dyn}}$ can be calculated in both the $\hat{\sigma}^x$- and $\hat{\sigma}^z$-basis. In the $\hat{\sigma}^x$-basis, we use Eq.~(9) of \cite{Wang2015}.

\paragraph{\ttcode{staggered\_imaginary\_times}} Order parameter from \cite{Wu2012}. It is defined as 
\begin{align}
    O_{\mathrm{SI}}^{x/z} = &\frac{1}{\beta} \bigl[ ( \tau_1^k - 0 ) - ( \tau_2^k - \tau_1^k ) + ... + (-1)^{N(k)-1} \nonumber \\
    &( \tau_{N(k)}^k - \tau_{N(k)-1}^k ) + (-1)^{N(k)} ( \beta - \tau_{N(k)}^k ) \bigr],
\end{align}
where $\tau_n^k$ is the imaginary time of the $n$-th tuple spin flip of type $k$. $k$ is a plaquette $p$ (star $s$) of links in the $\hat{\sigma}^x$-basis ($\hat{\sigma}^z$-basis). This order parameter can neither be evaluated from snapshots, nor from any other method that does not have access to imaginary time.

\paragraph{\ttcode{star\_x}} The star expectation value $\langle \hat{A}_v \rangle$.

\paragraph{\ttcode{string\_number}} The total number of links with $\hat{\sigma}^x=-1$ in the $\hat{\sigma}^x$-basis ($\hat{\sigma}^z=-1$ in the $\hat{\sigma}^z$-basis).

\section{Installation \& interfaces}
\label{sec:using}
There are five ways to use ParaToric, directly within code (C, C++, Python) or via the command-line (C++, Python). All interfaces require compiling C++ code. We tested the compilation with GCC 15 and Clang 20. 

All interfaces implement three functionalities. \textit{Thermalization} simulations are used to benchmark the thermalization process of a Markov chain and are primarily a diagnostic tool. Regular \textit{sampling} routines are used for generating snapshots and measuring observables, e.g., in the context of continuous phase transitions. \textit{Hysteresis} routines are a variant of the regular sampling routines where not one but an array of Hamiltonian parameters is provided and only one Markov chain is used for all parameters. The order of the Hamiltonian parameters in the input array matters: The last state of the previous parameter is used as an initial state for the thermalization phase of the next parameters. This simulation type should primarily be used when mapping out hysteresis curves in the vicinity of first-order phase transitions, hence the name. Since the hysteresis simulation returns the values of not one but many parameter sets, the output types are generally different from the regular sampling. It is also much slower than regular sampling, because the simulation for different parameters can in general not be parallelized.

\subsection{C++ interface}
The C++ interface enables users to use a ParaToric public header from within another C++ project. 

\subsubsection{Build \& Installation}

The core requires C++23, \href{https://cmake.org/}{CMake} $\geq$3.23 and \href{https://www.boost.org/}{Boost} $\geq$1.87 (older Boost versions may work, but were not tested). To compile it, run:

\begin{verbatim}
cmake -S . -B build -DCMAKE_BUILD_TYPE=Release \
-DPARATORIC_ENABLE_NATIVE_OPT=ON -DPARATORIC_LINK_MPI=OFF \
-DPARATORIC_BUILD_TESTS=ON
cmake --build build -jN
ctest --test-dir build -jN --output-on-failure
cmake --install build
\end{verbatim}

Replace \ttcode{N} with the number of cores to use, e.g.\ \ttcode{-j4} for 4 cores.

\medskip
\begin{itemize}
    \item \ttcode{-DCMAKE\_BUILD\_TYPE=Release}. Only set to \ttcode{Debug} if you're a developer.
    \item \ttcode{-DCMAKE\_INSTALL\_PREFIX}. By default, executables install to \ttcode{\$\{CMAKE\_SOURCE\_DIR\}/\$\{CMAKE\_INSTALL\_BINDIR\}/}, headers to \ttcode{\$\{CMAKE\_INSTALL\_INCLUDEDIR\}/paratoric}, and static libraries to \ttcode{\$\{CMAKE\_SOURCE\_DIR\}/\$\{CMAKE\_INSTALL\_LIBDIR\}/}. The Python scripts expect \ttcode{\$\{CMAKE\_SOURCE\_DIR\}/bin/}; this directory always contains the \ttcode{paratoric} executable. To install into a custom directory, pass it via \ttcode{-DCMAKE\_INSTALL\_PREFIX}, e.g. \ttcode{-DCMAKE\_INSTALL\_PREFIX=/your/custom/directory/}.
    \item \ttcode{-DPARATORIC\_EXPORT\_COMPILE\_COMMANDS=ON}. Export \ttcode{compile\_commands.json} for tooling.
    \item \ttcode{-DPARATORIC\_LINK\_MPI=OFF}. Link the core to MPI, which is required on some clusters. The core itself does not need MPI.
    \item \ttcode{-DPARATORIC\_ENABLE\_NATIVE\_OPT=OFF}. Use \ttcode{-march=native} on GCC and Clang.
    \item \ttcode{-DPARATORIC\_ENABLE\_AVX2=OFF}. Use AVX2 (Haswell New Instructions). Requires a CPU which supports AVX2.
    \item \ttcode{-DPARATORIC\_ENABLE\_FAST\_MATH=ON}. Use \ttcode{-ffast-math} on GCC and Clang.
    \item \ttcode{-DPARATORIC\_BUILD\_TESTS=PROJECT\_IS\_TOP\_LEVEL}. Compile the tests (recommended).
\end{itemize}

\subsubsection*{CMake usage (installed package)}
\begin{verbatim}
cmake_minimum_required(VERSION 3.23)
project(my_qmc_app CXX)

find_package(paratoric CONFIG REQUIRED)   # provides paratoric::core

add_executable(myapp main.cpp)
target_link_libraries(myapp PRIVATE paratoric::core)
\end{verbatim}

\subsubsection*{CMake usage (as subdirectory)}
If the core lives in \ttcode{deps/paratoric}, add it and link to the same target:
\begin{verbatim}
add_subdirectory(deps/paratoric)
add_executable(myapp main.cpp)
target_link_libraries(myapp PRIVATE paratoric::core)
\end{verbatim}

\subsubsection{Public class \ttcode{ExtendedToricCode}}
The interface class \ttcode{ExtendedToricCode} lives in the public header \ttcode{\#include <paratoric/mcmc/extended\_toric\_code.hpp>}. All symbols are in the \ttcode{paratoric} namespace. All methods are static, take a single \ttcode{Config} object and return a \ttcode{Result} object. The required fields in config are documented for each method within the docstrings.

\paragraph{\ttcode{Result \ ExtendedToricCode::get\_thermalization(Config \ config)}}
Run thermalization only.  
Required fields:  
\ttcode{lat\_spec.\{basis,lattice\_type,system\_size,beta,boundaries,default\_spin\}},  
\ttcode{param\_spec.\{mu,h,J,lmbda,h\_therm,lmbda\_therm\}},  
\ttcode{sim\_spec.\{N\_thermalization,N\_resamples,custom\_therm,observables,seed\}},  
\ttcode{out\_spec.\{path\_out,save\_snapshots\}}.

\paragraph{\ttcode{Result \ ExtendedToricCode::get\_sample(Config \ config)}}
Run a production measurement pass. Returns the observables selected in \ttcode{config}.  
Required fields:  
\ttcode{lat\_spec.\{basis,lattice\_type,system\_size,beta,boundaries,default\_spin\}},  
\ttcode{param\_spec.\{mu,h,J,lmbda\}},  
\ttcode{sim\_spec.\{N\_samples,N\_thermalization,N\_between\_samples,N\_resamples,observables,seed\}},  
\ttcode{out\_spec.\{path\_out,save\_snapshots\}}.

\paragraph{\ttcode{Result \ ExtendedToricCode::get\_hysteresis(Config \ config)}} 
Perform a hysteresis sweep, where the last state of the previous parameter is used as the initial state of the following parameter in \ttcode{h\_hys} \& \ttcode{lmbda\_hys}.  
Required fields:  
\ttcode{lat\_spec.\{basis,lattice\_type,system\_size,beta,boundaries,default\_spin\}},  
\ttcode{param\_spec.\{mu,J,h\_hys,lmbda\_hys\}},  
\ttcode{sim\_spec.\{N\_samples,N\_thermalization,N\_between\_samples,N\_resamples,observables,seed\}},  
\ttcode{out\_spec.\{paths\_out,save\_snapshots\}}.

\subsubsection{Configuration type}
\label{subsec:config}
The struct \ttcode{Config} (declared in \ttcode{<paratoric/types/types.hpp>}) contains multiple nested specifications.

\subsubsection*{Top-level configuration: \ttcode{Config}}
\begin{center}
\begin{tabular}{lll}
\hline
\textbf{Field} & \textbf{Type} & \textbf{Purpose}\\
\hline
\ttcode{sim\_spec}   & \ttcode{SimSpec}                & Simulation / MC controls (backend-consumed).\\
\ttcode{param\_spec} & \ttcode{ParamSpec} & Model couplings / parameters (backend-consumed).\\
\ttcode{lat\_spec} & \ttcode{LatSpec}                 & Lattice geometry and basis.\\
\ttcode{out\_spec}   & \ttcode{OutSpec}                    & Output folders and snapshot toggles.\\
\hline
\end{tabular}
\end{center}

\subsubsection*{Simulation specification (\ttcode{config.sim\_spec})}
\begin{center}
\begin{tabular}{llp{0.52\linewidth}}
\hline
\textbf{Field} & \textbf{Type} & \textbf{Meaning / Defaults}\\
\hline
\ttcode{N\_samples} & \ttcode{int} & Number of recorded snapshots. Default \ttcode{1000}.\\
\ttcode{N\_thermalization} & \ttcode{int} & Number of warmup steps before sampling. Typically $O(L^d)$, where $L$ is the system size and $d$ is the dimensionality. Default \ttcode{10000}.\\
\ttcode{N\_between\_samples} & \ttcode{int} & Steps between consecutive snapshots. Higher value decreases autocorrelation and improves error bars. Typically $O(L^d)$, where $L$ is the system size and $d$ is the dimensionality. Default \ttcode{1000}.\\
\ttcode{N\_resamples} & \ttcode{int} & Bootstrap resamples for errors. Default \ttcode{1000}.\\
\ttcode{custom\_therm} & \ttcode{bool} & Use custom thermalization schedule. Default \mbox{\ttcode{false}}.\\
\ttcode{seed} & \ttcode{int} & PRNG seed. \ttcode{0} means “random seed.” Default \ttcode{0}.\\
\ttcode{observables} & \ttcode{vector<string>} & Names of observables to record each snapshot. For options, see Sec.~\ref{sec:observables}.\\
\hline
\end{tabular}
\end{center}

\subsubsection*{Parameter specification (\ttcode{config.param\_spec})}
\begin{center}
\begin{tabular}{llp{0.52\linewidth}}
\hline
\textbf{Field} & \textbf{Type} & \textbf{Meaning / Defaults}\\
\hline
\ttcode{mu} & \ttcode{double} & Star term coefficient. Default \ttcode{1.0}.\\
\ttcode{h} & \ttcode{double} & Electric field term. Default \ttcode{0.0}.\\
\ttcode{J} & \ttcode{double} & Plaquette term. Default \ttcode{1.0}.\\
\ttcode{lmbda} & \ttcode{double} & Gauge-field term. Default \ttcode{0.0}.\\
\ttcode{h\_therm} & \ttcode{double} & Thermalization value for \ttcode{h} when using custom schedules. Default \ttcode{NaN} (unused).\\
\ttcode{lmbda\_therm} & \ttcode{double} & Thermalization value for \ttcode{lmbda} when using custom schedules. Default \ttcode{NaN} (unused).\\
\ttcode{h\_hys} & \ttcode{vector<double>} & Sweep values of \ttcode{h} for hysteresis runs. Default empty. Length must match \ttcode{lmbda\_hys}.\\
\ttcode{lmbda\_hys} & \ttcode{vector<double>} & Sweep values of \ttcode{lmbda} for hysteresis runs. Default empty. Length must match \ttcode{h\_hys}.\\
\hline
\end{tabular}
\end{center}

\subsubsection*{Lattice specification (\ttcode{config.lat\_spec})}

\begin{center}
\begin{tabular}{llp{0.52\linewidth}}
\hline
\textbf{Field} & \textbf{Type} & \textbf{Meaning / Valid values}\\
\hline
\ttcode{basis} & \ttcode{char} & Spin eigenbasis for the simulation. Must be \ttcode{'x'} or \ttcode{'z'}.\\
\ttcode{lattice\_type} & \ttcode{string} & The lattice (\ttcode{"square"}, \ttcode{"triangular"}, \mbox{\ttcode{"honeycomb"}} or \ttcode{"cubic"}).\\
\ttcode{system\_size} & \ttcode{int} & Linear system size (per dimension).\\
\ttcode{beta} & \ttcode{double} & Inverse temperature $\beta>0$.\\
\ttcode{boundaries} & \ttcode{string} & Boundary condition: \ttcode{"periodic"} or \ttcode{"open"}.\\
\ttcode{default\_spin} & \ttcode{int} & Initial link spin, must be $+1$ or $-1$.\\
\hline
\end{tabular}
\end{center}

\subsubsection*{Output specification (\ttcode{config.out\_spec})}
\begin{center}
\begin{tabular}{lll}
\hline
\textbf{Field} & \textbf{Type} & \textbf{Meaning}\\
\hline
\ttcode{path\_out}   & \ttcode{string}           & Primary output folder name.\\
\ttcode{paths\_out}  & \ttcode{vector<string>} & Hysteresis subfolder names. Length must match \ttcode{h\_hys}.\\
\ttcode{save\_snapshots} & \ttcode{bool}                  & Save snapshots toggle. Default \ttcode{false}.\\
\hline
\end{tabular}
\end{center}

\subsubsection{Return type}
\label{subsec:results}

\begin{center}
\begin{tabularx}{\linewidth}{F T M}
\hline
\textbf{Field} & \multicolumn{1}{c}{\textbf{C++ Type}} & \textbf{Meaning}\\
\hline
\ttcode{series} & \tcode{vector<vector<variant<complex<double>, double>>>} &
Time series of all requested observables, thermalization is excluded (except for thermalization simulation). Outer index = observable, inner index = time point.\\
\ttcode{acc\_ratio} & \tcode{vector<double>} & Time series of Monte Carlo acceptance ratios.\\
\ttcode{mean}   & \tcode{vector<double>} & Bootstrap observable means.\\
\ttcode{mean\_std} & \tcode{vector<double>} & Bootstrap standard errors of the mean.\\
\ttcode{binder} & \tcode{vector<double>} & Bootstrap binder ratios.\\
\ttcode{binder\_std} & \tcode{vector<double>} & Bootstrap standard errors of the binder ratios.\\
\ttcode{tau\_int} & \tcode{vector<double>} & Estimated integrated autocorrelation times.\\[0.2em]
\ttcode{series\_hys} & \tcode{vector<vector<vector<variant<complex<double>, double>>>>} & Hysteresis time series of all requested observables, thermalization is excluded (except for thermalization simulation). Outer vector = hysteresis parameters (order as in \ttcode{h\_hys}, \mbox{\ttcode{lmbda\_hys}}), middle vector = observables (order as in \mbox{\ttcode{observables}}), inner vector = time series.\\
\ttcode{mean\_hys}      & \tcode{vector<vector<double>>} & Hysteresis bootstrap observable means. Outer vector = hysteresis parameters (order as in \ttcode{h\_hys}, \mbox{\ttcode{lmbda\_hys}}), inner vector = observables (order as in \mbox{\ttcode{observables}}).\\
\ttcode{mean\_std\_hys} & \tcode{vector<vector<double>>} & Hysteresis bootstrap standard errors of the mean. Indices as above.\\
\ttcode{binder\_hys}    & \tcode{vector<vector<double>>} & Hysteresis bootstrap binder ratios. Indices as above.\\
\ttcode{binder\_std\_hys} & \tcode{vector<vector<double>>} & Hysteresis bootstrap standard errors of the binder ratios. Indices as above.\\
\ttcode{tau\_int\_hys}  & \tcode{vector<vector<double>>} & Hysteresis estimated integrated autocorrelation times. Indices as above.\\
\hline
\end{tabularx}
\end{center}

\newpage

\subsubsection{C++ usage examples}

\lstset{
  language=C++,
  basicstyle=\ttfamily\small,
  keywordstyle=\bfseries,
  columns=fullflexible,
  upquote=true,
  frame=single,
  breaklines=true
}
\begin{lstlisting}[language=C++,basicstyle=\ttfamily\small\color{darkblue},caption={C++ API - Minimal call}]
// C++23
#include <iostream>
#include <print>
#include <vector>
#include <string>
#include <paratoric/mcmc/extended_toric_code.hpp>
#include <paratoric/types/types.hpp>

int main() {
  using namespace paratoric;

  Config cfg{};

  // ---- lattice sub-config (required) ----
  cfg.lat_spec.basis        = 'z';          // or 'x'
  cfg.lat_spec.lattice_type = "square";     // or "cubic", "honeycomb", ...
  cfg.lat_spec.system_size  = 16;
  cfg.lat_spec.beta         = 8.0;
  cfg.lat_spec.boundaries   = "periodic";   // or "open"
  cfg.lat_spec.default_spin = 1;

  // ---- Hamiltonian parameters ----
  cfg.param_spec.mu    = 1.0;    // star term
  cfg.param_spec.J     = 1.0;    // plaquette term
  cfg.param_spec.h     = 0.20;   // electric field term
  cfg.param_spec.lmbda = 0.00;   // gauge-field term

  // Optional thermalization schedule values (used if custom_therm = true)
  cfg.param_spec.h_therm     = std::numeric_limits<double>::quiet_NaN();
  cfg.param_spec.lmbda_therm = std::numeric_limits<double>::quiet_NaN();

  // (Optional) Hysteresis sweep grids - only read by get_hysteresis(...)
  cfg.param_spec.h_hys     = {};            // e.g. {0.0, 0.1, 0.2, 0.3, 0.2, 0.1, 0.0}
  cfg.param_spec.lmbda_hys = {};            // e.g. {0.0, 0.1, 0.2, 0.3, 0.2, 0.1, 0.0}

  // ---- Simulation (MC) controls ----
  cfg.sim_spec.N_samples         = 0;       // 0 => thermalization-only
  cfg.sim_spec.N_thermalization  = 5000;    // warmup steps
  cfg.sim_spec.N_between_samples = 10;      // thinning between snapshots
  cfg.sim_spec.N_resamples       = 1000;    // bootstrap
  cfg.sim_spec.custom_therm      = false;   // set true to use *_therm values
  cfg.sim_spec.seed              = 12345;   // 0 => random seed

  // Observables to record each snapshot (backend-recognized names)
  cfg.sim_spec.observables = {
    "energy",           // total energy
    "plaquette_z",      // plaquette energy 
    "anyon_count",           // number of anyons (x-basis: e-anyons, z-basis: m-anyons)
    "fredenhagen_marcu"   // example: Wilson/'t Hooft loop proxy
  };

  // ---- Output / I/O policy ----
  cfg.out_spec.path_out                 = "runs/sample";  // single-run output dir
  cfg.out_spec.paths_out                = {};             // filled only for hysteresis
  cfg.out_spec.save_snapshots           = false;          // set true to dump every snapshot
  cfg.out_spec.full_time_series = true;           // save full time series (FCS)

  // 1) Check thermalization
  Result warmup = ExtendedToricCode::get_thermalization(cfg);
  std::print("Thermalization series: {}", warmup.series);

  // 2) Production sample (set N_samples > 0 and call get_sample)
  cfg.sim_spec.N_samples = 2000;
  Result out = ExtendedToricCode::get_sample(cfg);
  std::print("Production autocorrelations: {}", out.tau_int);

  return 0;
}

\end{lstlisting}

\subsection{C++ command-line interface}
ParaToric ships a C++ command-line interface \ttcode{\$\{CMAKE\_INSTALL\_PREFIX\}/\$\{CMAKE\_INSTALL\_BINDIR\}/paratoric} that orchestrates C\texttt{++} backends, runs sweeps, and writes HDF5 (observables) and XML (snapshots) outputs.

\subsubsection{Build \& Installation}

The command-line interface requires \href{https://www.hdfgroup.org/}{HDF5} $\geq$1.14.3 (older HDF5 versions may work, but were not tested). The core requires C++23, \href{https://cmake.org/}{CMake} $\geq$3.23 and \href{https://www.boost.org/}{Boost} $\geq$1.87 (older Boost versions may work, but were not tested). To compile it, run:

\begin{verbatim}
cmake -S . -B build -DCMAKE_BUILD_TYPE=Release \
-DPARATORIC_ENABLE_NATIVE_OPT=ON -DPARATORIC_LINK_MPI=OFF \
-DPARATORIC_BUILD_TESTS=ON -DPARATORIC_BUILD_CLI=ON
cmake --build build -jN
ctest --test-dir build -jN --output-on-failure
cmake --install build
\end{verbatim}

Replace \ttcode{N} with the number of cores to use, e.g.\ \ttcode{-j4} for 4 cores.

\medskip
\begin{itemize}
    \item \ttcode{-DCMAKE\_BUILD\_TYPE=Release}. Only set to \ttcode{Debug} if you're a developer.
    \item \ttcode{-DCMAKE\_INSTALL\_PREFIX}. By default, executables install to \ttcode{\$\{CMAKE\_SOURCE\_DIR\}/\$\{CMAKE\_INSTALL\_BINDIR\}/}, headers to \ttcode{\$\{CMAKE\_INSTALL\_INCLUDEDIR\}/paratoric}, and static libraries to \ttcode{\$\{CMAKE\_SOURCE\_DIR\}/\$\{CMAKE\_INSTALL\_LIBDIR\}/}. The Python scripts expect \ttcode{\$\{CMAKE\_SOURCE\_DIR\}/bin/}; this directory always contains the \ttcode{paratoric} executable. To install into a custom directory, pass it via \ttcode{-DCMAKE\_INSTALL\_PREFIX}, e.g. \ttcode{-DCMAKE\_INSTALL\_PREFIX=/your/custom/directory/}.
    \item \ttcode{-DPARATORIC\_EXPORT\_COMPILE\_COMMANDS=ON}. Export \ttcode{compile\_commands.json} for tooling.
    \item \ttcode{-DPARATORIC\_LINK\_MPI=OFF}. Link the core to MPI, which is required on some clusters. The core itself does not need MPI.
    \item \ttcode{-DPARATORIC\_ENABLE\_NATIVE\_OPT=OFF}. Use \ttcode{-march=native} on GCC and Clang.
    \item \ttcode{-DPARATORIC\_ENABLE\_AVX2=OFF}. Use AVX2 (Haswell New Instructions). Requires a CPU which supports AVX2.
    \item \ttcode{-DPARATORIC\_ENABLE\_FAST\_MATH=ON}. Use \ttcode{-ffast-math} on GCC and Clang.
    \item \ttcode{-DPARATORIC\_BUILD\_TESTS=PROJECT\_IS\_TOP\_LEVEL}. Compile the tests (recommended).
    \item \ttcode{-DPARATORIC\_BUILD\_CLI=PROJECT\_IS\_TOP\_LEVEL} Required for both the C++ and Python command line interface.
\end{itemize}

\subsubsection*{Global options}

\begin{center}
\small
\begin{tabularx}{\linewidth}{
  >{\raggedright\arraybackslash}p{0.3\linewidth}
  >{\raggedright\arraybackslash}p{0.16\linewidth}
  >{\raggedright\arraybackslash}p{0.14\linewidth}
  X
}
\hline
\textbf{Long flag} & \textbf{Short} & \textbf{Type} & \textbf{Description}\\
\hline
\ttcode{--simulation} & \ttcode{-sim} & string & Simulation mode: \mbox{\ttcode{etc\_sample}}, \mbox{\ttcode{etc\_hysteresis}}, \mbox{\ttcode{etc\_thermalization}}.\\
\ttcode{--N\_samples} & \ttcode{-Ns} & int & Number of recorded samples.\\
\ttcode{--N\_thermalization} & \ttcode{-Nth} & int & Thermalization (warmup) steps.\\
\ttcode{--N\_between\_samples} & \ttcode{-Nbs} & int & Steps between samples (thinning).\\
\ttcode{--beta} & \ttcode{-bet} & double & Inverse temperature $\beta=1/T$.\\
\ttcode{--mu\_constant} & \ttcode{-muc} & double & Star-term coupling $\mu$.\\
\ttcode{--J\_constant} & \ttcode{-Jc} & double & Plaquette coupling $J$.\\
\ttcode{--h\_constant} & \ttcode{-hc} & double & Field $h$.\\
\ttcode{--lmbda\_constant} & \ttcode{-lmbdac} & double & Field $\lambda$.\\
\ttcode{--h\_constant\_therm} & \ttcode{-hct} & double & Thermalization value for $h$ (used if custom therm).\\
\ttcode{--lmbda\_constant\_therm} & \ttcode{-lmbdact} & double & Thermalization value for $\lambda$.\\
\ttcode{--h\_hysteresis} & \ttcode{-hhys} & list<double> & Hysteresis schedule for $h$ (space-separated). Length must match \mbox{\ttcode{lmbdahys}}.\\
\ttcode{--lmbda\_hysteresis} & \ttcode{-lmbdahys} & list<double> & Hysteresis schedule for $\lambda$. Length must match \ttcode{hhys}.\\
\ttcode{--N\_resamples} & \ttcode{-Nr} & int & Bootstrap resamples (error bars).\\
\ttcode{--custom\_therm} & \ttcode{-cth} & bool & Use thermalization values (\ttcode{0}/\ttcode{1}).\\
\ttcode{--observables} & \ttcode{-obs} & list<string> & Measured observables (space-separated).\\
\ttcode{--seed} & \ttcode{-s} & int & PRNG seed; \ttcode{0} means random seed.\\
\ttcode{--basis} & \ttcode{-bas} & char & Spin basis: \ttcode{'x'} or \ttcode{'z'}.\\
\ttcode{--lattice\_type} & \ttcode{-lat} & string & Lattice type (e.g.\ \mbox{\ttcode{square}}, \mbox{\ttcode{cubic}}, \dots).\\
\ttcode{--system\_size} & \ttcode{-L} & int & Linear lattice size (per dimension).\\
\ttcode{--boundaries} & \ttcode{-bound} & string & \ttcode{periodic} or \ttcode{open}.\\
\ttcode{--default\_spin} & \ttcode{-dsp} & int & Initial link spin (\ttcode{+1} or \ttcode{-1}).\\
\ttcode{--output\_directory} & \ttcode{-outdir} & path & Output directory path.\\
\ttcode{--folder\_name} & \ttcode{-fn} & string & Subfolder (of output directory) name for single run.\\
\ttcode{--folder\_names} & \ttcode{-fns} & list<string> & Subfolders (of output directory) for hysteresis steps. Length must match \ttcode{lmbdahys}.\\
\ttcode{--snapshots} & \ttcode{-snap} & bool & Save snapshots into specified subfolders of output directory.\\
\ttcode{--full\_time\_series} & \ttcode{-fts} & bool & Save full time series toggle.\\
\ttcode{--process\_index} & \ttcode{-procid} & int & Process identifier (logging/debug).\\
\hline
\end{tabularx}
\end{center}

\newpage

\subsubsection{\ttcode{etc\_sample}}
Runs a production measurement pass with the supplied configuration.

\begin{lstlisting}[language=bash, basicstyle=\ttfamily\small\color{darkblue}, breaklines=true, caption={Example usage}]
./paratoric -sim etc_sample -Ns 2000 -Nth 5000 -Nbs 10 -Nr 1000 -bet 16.0 -muc 1 -Jc 1 -hc 0.2 -lmbdac 0.0 -obs energy plaquette_z anyon_count -bas z -lat square -L 16 -bound periodic -dsp 1 -outdir ./runs/sample -snap=0 -fts=1
\end{lstlisting}

\subsubsection{\ttcode{etc\_hysteresis}}
Runs a parameter sweep where the last state of step $i$ initializes step $i{+}1$.%
\ Provide \ttcode{--h\_hysteresis} and \ttcode{--lmbda\_hysteresis} as space-separated lists, and \ttcode{--folder\_names} for per-step outputs.

\begin{lstlisting}[language=bash, basicstyle=\ttfamily\small\color{darkblue}, breaklines=true, caption={Example usage}]
./paratoric -sim etc_hysteresis -Ns 1000 -Nth 2000 -Nbs 50 -Nr 500 -bet 12.0 -muc 1 -Jc 1 -lmbdahys 0.2 0.2 0.2 0.2 0.2 0.2 0.2 -hhys 0.0 0.1 0.2 0.3 0.2 0.1 0.0 -obs energy fredenhagen_marcu -bas x -lat square -L 12 -bound periodic -dsp 1 -outdir ./runs/hys -fns step0 step1 step2 step3 step4 step5 step6
\end{lstlisting}

\subsubsection{\ttcode{etc\_thermalization}}
Performs thermalization only (no production sampling).

\begin{lstlisting}[language=bash, basicstyle=\ttfamily\small\color{darkblue}, breaklines=true, caption={Example usage}]
./paratoric -sim etc_thermalization -Ns 0 -Nth 5000 -Nbs 10 -Nr 500 -bet 10.0 -muc 1 -Jc 1 -hc 0.3 -lmbdac 0.1 -hct 0.4 -lmbdact 0.2 -cth 1 -obs energy anyon_density -bas z -lat square -L 10 -bound open -dsp 1 -outdir ./runs/therm -snap=1
\end{lstlisting}

\subsubsection{HDF5 structure}

The output HDF5 has the structure \ttcode{simulation/results/acc\_ratio} for an array of the acceptance ratios (only for thermalization), \ttcode{simulation/results/observable\_name/series} for the time series (if it was enabled), and \ttcode{simulation/results/observable\_name/\{mean, mean\_error, binder, binder\_error, autocorrelation\_time\}}. For the regular sampling, \ttcode{mean}, \ttcode{mean\_error}, \ttcode{binder}, \ttcode{binder\_error} and \ttcode{autocorrelation\_time} contain doubles. For the hysteresis, they contain an array of values for the hysteresis parameters (in the order of \ttcode{h\_hysteresis} and \ttcode{lmbda\_hysteresis}).

\subsection{C interface}
The C interface enables users to use a ParaToric public header from within another C project or another programming language that supports a C-style interface.  The C interface exposes a stable ABI. It mirrors the C\texttt{++} interface. Include the public header \ttcode{\#include <paratoric/mcmc/extended\_toric\_code\_c.h>}.  
All functions return \ttcode{ptc\_status\_t}.

\subsubsection{Build \& Installation}

The code can be compiled in exactly the same fashion as for the C++ interface.

\paragraph{CMake usage (installed package).}
\begin{verbatim}
cmake_minimum_required(VERSION 3.23)
project(my_qmc_c C)
find_package(paratoric CONFIG REQUIRED)   # provides paratoric::core
add_executable(cdemo main.c)
target_link_libraries(cdemo PRIVATE paratoric::core)
\end{verbatim}

\paragraph{CMake usage (as subdirectory).}
\begin{verbatim}
add_subdirectory(deps/paratoric)
add_executable(cdemo main.c)
target_link_libraries(cdemo PRIVATE paratoric::core)
\end{verbatim}

\subsubsection{Status \& error handling}
\begin{center}
\begin{tabularx}{\linewidth}{l X X}
\hline
\textbf{Name} & \textbf{C type/values} & \textbf{Meaning}\\
\hline
\ttcode{ptc\_status\_t} &
\tcode{\{PTC\_STATUS\_OK=0,\ PTC\_STATUS\_INVALID\_ARGUMENT=1,\ PTC\_STATUS\_RUNTIME\_ERROR=2,\ PTC\_STATUS\_NO\_MEMORY=3,\ PTC\_STATUS\_INTERNAL\_ERROR=4\}} &
Return code of every API call.\\
\ttcode{ptc\_last\_error()} & \ttcode{const\ char*} & Thread–local error string. Valid until next call.\\
\hline
\end{tabularx}
\end{center}

\subsubsection{Opaque handle}
Create and destroy the interface instance.  
Use \ttcode{ptc\_create(ptc\_handle\_t\ **out)} and \ttcode{ptc\_destroy(ptc\_handle\_t\ *h)}.

\subsubsection{Configuration type}
\label{subsec:c-config}
Top-level \ttcode{ptc\_config\_t} aggregates four nested specs.  
Field names mirror the C\texttt{++} \ttcode{Config}.

\subsubsection*{Top-level configuration: \ttcode{ptc\_config\_t}}
\begin{center}
\begin{tabularx}{\linewidth}{l l X}
\hline
\textbf{Field} & \textbf{Type} & \textbf{Purpose}\\
\hline
\ttcode{sim}    & \ttcode{ptc\_sim\_spec\_t}    & Monte Carlo parameters.\\
\ttcode{params} & \ttcode{ptc\_param\_spec\_t}  & Hamiltonian parameters.\\
\ttcode{lat}    & \ttcode{ptc\_lat\_spec\_t}    & Lattice parameters.\\
\ttcode{out}    & \ttcode{ptc\_out\_spec\_t}    & Output paths and snapshot toggle.\\
\hline
\end{tabularx}
\end{center}

\subsubsection*{Simulation specification (\ttcode{config.sim})}
\begin{center}
\begin{tabularx}{\linewidth}{l l X}
\hline
\textbf{Field} & \textbf{Type} & \textbf{Meaning}\\
\hline
\ttcode{N\_samples}          & \ttcode{int}                 & Number of snapshots.\\
\ttcode{N\_thermalization}   & \ttcode{int}                 & Thermalization steps.\\
\ttcode{N\_between\_samples} & \ttcode{int}                 & Thinning between snapshots.\\
\ttcode{N\_resamples}        & \ttcode{int}                 & Bootstrap resamples.\\
\ttcode{custom\_therm}       & \ttcode{bool}                & Custom thermalization schedule.\\
\ttcode{seed}                & \ttcode{int}                 & PRNG seed (\ttcode{0} = random).\\
\ttcode{observables}         & \ttcode{const\ char*\ const*}& Array of observable names (nullable).\\
\ttcode{N\_observables}      & \ttcode{size\_t}             & Length of \ttcode{observables}.\\
\hline
\end{tabularx}
\end{center}

\subsubsection*{Parameter specification (\ttcode{config.params})}
\begin{center}
\begin{tabularx}{\linewidth}{l l X}
\hline
\textbf{Field} & \textbf{Type} & \textbf{Meaning}\\
\hline
\ttcode{mu,h,J,lmbda}     & \ttcode{double}          & Couplings (star, electric, plaquette, gauge).\\
\ttcode{h\_therm}         & \ttcode{double}          & Thermalization value for \ttcode{h} if \ttcode{custom\_therm=true}.\\
\ttcode{lmbda\_therm}     & \ttcode{double}          & Thermalization value for \ttcode{lmbda} if \ttcode{custom\_therm=}\mbox{\ttcode{true}}.\\
\ttcode{h\_hys}           & \ttcode{const\ double*}  & Hysteresis schedule for \ttcode{h} (nullable).\\
\ttcode{h\_hys\_len}      & \ttcode{size\_t}         & Length of \ttcode{h\_hys}. Must match \ttcode{lmbda\_hys\_len}.\\
\ttcode{lmbda\_hys}       & \ttcode{const\ double*}  & Hysteresis schedule for \ttcode{lmbda} (nullable).\\
\ttcode{lmbda\_hys\_len}  & \ttcode{size\_t}         & Length of \ttcode{lmbda\_hys}. Must match \ttcode{h\_hys\_len}.\\
\hline
\end{tabularx}
\end{center}

\subsubsection*{Lattice specification (\ttcode{config.lat})}
\begin{center}
\begin{tabularx}{\linewidth}{l l X}
\hline
\textbf{Field} & \textbf{Type} & \textbf{Meaning / Valid values}\\
\hline
\ttcode{basis}        & \ttcode{char}           & Spin basis: \ttcode{'x'} or \ttcode{'z'}.\\
\ttcode{lattice\_type} & \ttcode{const\ char*}   & E.g.\ \ttcode{"triangular"}, \ttcode{"square"}, \ldots\\
\ttcode{system\_size}  & \ttcode{int}            & Linear system size per dimension.\\
\ttcode{beta}          & \ttcode{double}         & Inverse temperature.\\
\ttcode{boundaries}    & \ttcode{const\ char*}   & \ttcode{"periodic"} or \ttcode{"open"}.\\
\ttcode{default\_spin} & \ttcode{int}            & Initial link spin: \ttcode{+1} or \ttcode{-1}.\\
\hline
\end{tabularx}
\end{center}

\subsubsection*{Output specification (\ttcode{config.out})}
\begin{center}
\begin{tabularx}{\linewidth}{l l X}
\hline
\textbf{Field} & \textbf{Type} & \textbf{Meaning}\\
\hline
\ttcode{path\_out}     & \ttcode{const\ char*}        & Single output directory (nullable).\\
\ttcode{paths\_out}    & \ttcode{const\ char*\ const*}& Output directories for hysteresis steps (nullable).\\
\ttcode{N\_paths\_out} & \ttcode{size\_t}             & Length of \ttcode{paths\_out}. Must match \mbox{\ttcode{h\_hys\_len}} and \ttcode{lmbda\_hys\_len}. \\
\ttcode{save\_snapshots} & \ttcode{bool}              & Toggle snapshot dumping.\\
\hline
\end{tabularx}
\end{center}

\newpage

\subsubsection{Return type}
All outputs are owned by the caller.  
Call \ttcode{ptc\_result\_destroy(\&r)} to free and zero.

\begin{center}
\small
\begin{tabularx}{\linewidth}{
  >{\raggedright\arraybackslash}p{0.38\linewidth}
  >{\raggedright\arraybackslash}p{0.22\linewidth}
  X                                           
}
\hline
\textbf{Field} & \textbf{C type} & \textbf{Meaning}\\
\hline
\ttcode{series} & \ttcode{ptc\_series\_t} & Time series (real/complex) of all requested observables, thermalization is excluded (except for thermalization simulation). Outer index = observable, inner index = time point.\\
\ttcode{acc\_ratio} & \ttcode{ptc\_dvec\_t} & MC acceptance ratios for each update (thermalization).\\
\ttcode{mean}, \ttcode{mean\_std} & \ttcode{ptc\_dvec\_t} & Bootstrap mean and standard error (order as in \ttcode{observables}).\\
\ttcode{binder}, \ttcode{binder\_std} & \ttcode{ptc\_dvec\_t} & Binder ratios and standard error (order as in \ttcode{observables}).\\
\ttcode{tau\_int} & \ttcode{ptc\_dvec\_t} & Integrated autocorrelation time (order as in \mbox{\ttcode{observables}}).\\
\ttcode{series\_hys} & \ttcode{ptc\_series\_blocks\_t} & Hysteresis time series. Outer index = hysteresis parameters (order as in \ttcode{h\_hys}, \mbox{\ttcode{lmbda\_hys}}), middle index = observables (order as in \mbox{\ttcode{observables}}), inner vector = time series.\\
\ttcode{mean\_hys}, \ttcode{mean\_std\_hys}, \ttcode{binder\_hys}, \ttcode{binder\_std\_hys}, \ttcode{tau\_int\_hys} & \ttcode{ptc\_dmat\_t} &  Outer index = hysteresis parameters (order as in \mbox{\ttcode{h\_hys}}, \mbox{\ttcode{lmbda\_hys}}), inner index = observables (order as in \mbox{\ttcode{observables}}).\\
\hline
\end{tabularx}
\end{center}

\subsubsection{Procedures (mirror the C\texttt{++} API)}
All fill a \ttcode{ptc\_result\_t\ *out} on success.  
Return \ttcode{PTC\_STATUS\_OK} on success.

\paragraph{\ttcode{ptc\_get\_thermalization(ptc\_handle\_t\ *h,\ const\ ptc\_config\_t\ *cfg,\ ptc\_result\_t\ *out)}}
Run thermalization only.  
\textbf{Required fields:}  
\ttcode{cfg->lat.\{basis,lattice\_type,system\_size,beta,boundaries,default\_spin\}},  
\ttcode{cfg->params.\{mu,h,J,lmbda\}},  
\ttcode{cfg->sim.\{N\_thermalization,N\_resamples,observables,N\_observables,seed\}},  
\ttcode{cfg->out.\{path\_out,save\_snapshots\}}.

\paragraph{\ttcode{ptc\_get\_sample(ptc\_handle\_t\ *h,\ const\ ptc\_config\_t\ *cfg,\ ptc\_result\_t\ *out)}}
Run a production measurement pass.  
\textbf{Required fields:}  
\ttcode{cfg->lat.\{basis,lattice\_type,system\_size,beta,boundaries,default\_spin\}},  
\ttcode{cfg->params.\{mu,h,J,lmbda,h\_therm,lmbda\_therm\}},  
\ttcode{cfg->sim.\{N\_samples,N\_thermalization,N\_between\_samples,N\_resamples,custom\_therm,observables,N\_observables,seed\}},  
\ttcode{cfg->out.\{path\_out,save\_snapshots\}}.

\paragraph{\ttcode{ptc\_get\_hysteresis(ptc\_handle\_t\ *h,\ const\ ptc\_config\_t\ *cfg,\ ptc\_result\_t\ *out)}}
Run a hysteresis sweep over \ttcode{h\_hys} and/or \ttcode{lmbda\_hys}.  
The last state of step \ttcode{i} initializes step \ttcode{i+1}.  
\textbf{Required fields:}  
\ttcode{cfg->lat.\{basis,lattice\_type,system\_size,beta,boundaries,default\_spin\}},  
\ttcode{cfg->params.\{mu,h\_hys,h\_hys\_len,J,lmbda\_hys,lmbda\_hys\_len\}},  
\ttcode{cfg->sim.\{N\_samples,N\_thermalization,N\_between\_samples,N\_resamples,observables,N\_observables,seed\}},  
\ttcode{cfg->out.\{paths\_out,N\_paths\_out,save\_snapshots\}}.

\subsubsection{C usage example}
\lstset{
  language=C,
  basicstyle=\ttfamily\small,
  columns=fullflexible,
  upquote=true,
  frame=single,
  breaklines=true
}
\begin{lstlisting}[language=C, basicstyle=\ttfamily\small\color{darkblue}, caption={C API - Minimal call}]
#include <stdio.h>
#include <math.h>
#include <paratoric/mcmc/extended_toric_code_c.h>

int main(void) {
  ptc_handle_t* h = NULL;
  if (ptc_create(&h) != PTC_STATUS_OK) { puts("create failed"); return 1; }

  ptc_lat_spec_t lat = {
    .basis = 'z',
    .lattice_type = "square",
    .system_size = 16,
    .beta = 8.0,
    .boundaries = "periodic",
    .default_spin = 1
  };

  ptc_param_spec_t ps = {
    .mu = 1.0, .h = 0.2, .J = 1.0, .lmbda = 0.0,
    .h_therm = NAN, .lmbda_therm = NAN,
    .h_hys = NULL, .h_hys_len = 0,
    .lmbda_hys = NULL, .lmbda_hys_len = 0
  };

  const char* obs[] = {"energy","plaquette_z","anyon_count"};
  ptc_sim_spec_t sim = {
    .N_samples = 0,                /* thermalization-only initially */
    .N_thermalization = 5000,
    .N_between_samples = 10,
    .N_resamples = 1000,
    .custom_therm = false,
    .seed = 12345,
    .observables = obs,
    .N_observables = sizeof(obs)/sizeof(obs[0])
  };

  ptc_out_spec_t outspec = {
    .path_out = "runs/sample",
    .paths_out = NULL, .N_paths_out = 0,
    .save_snapshots = false
  };

  ptc_config_t cfg = { .sim = sim, .params = ps, .lat = lat, .out = outspec };

  ptc_result_t warm = {0};
  ptc_status_t st = ptc_get_thermalization(h, &cfg, &warm);
  if (st != PTC_STATUS_OK) { puts(ptc_last_error()); ptc_destroy(h); return 2; }
  ptc_result_destroy(&warm);

  cfg.sim.N_samples = 2000;
  ptc_result_t res = {0};
  st = ptc_get_sample(h, &cfg, &res);
  if (st != PTC_STATUS_OK) { puts(ptc_last_error()); ptc_destroy(h); return 3; }

  // use res.mean, res.tau_int, ...
  ptc_result_destroy(&res);
  ptc_destroy(h);
  return 0;
}
\end{lstlisting}

\paragraph{Memory rules.}
You own all buffers in \ttcode{ptc\_result\_t}.  
Call \ttcode{ptc\_result\_destroy} once per successful call.

\subsection{Python bindings}
ParaToric exposes a compiled Python extension module \ttcode{\_paratoric} with a submodule \ttcode{extended\_toric\_code}.%
\ The bindings convert C\texttt{++} vectors into NumPy arrays and release the global interpreter lock (GIL) while running the C\texttt{++} kernels.

\subsubsection{Build \& Installation}

The core requires C++23, \href{https://cmake.org/}{CMake} $\geq$3.23 and \href{https://www.boost.org/}{Boost} $\geq$1.87 (older Boost versions may work, but were not tested). The Python bindings require a Python installation with Numpy \cite{Harris2020} and PyBind11 (tested with version 3.0.1). Pybind11 is included as a git submodule (you need to pull it!). To compile the Python bindings, run:

\begin{verbatim}
cmake -S . -B build -DCMAKE_BUILD_TYPE=Release \
-DPARATORIC_ENABLE_NATIVE_OPT=ON -DPARATORIC_LINK_MPI=OFF \
-DPARATORIC_BUILD_TESTS=ON -DPARATORIC_BUILD_PYBIND=ON \
-DPython3_EXECUTABLE="$(which python)" -DPYBIND11_FINDPYTHON=ON \
-DPARATORIC_INSTALL_TO_SITE=ON -DPARATORIC_PIP_EDITABLE_INSTALL=ON
cmake --build build -jN
ctest --test-dir build -jN --output-on-failure
cmake --install build
\end{verbatim}

Replace \ttcode{N} with the number of cores to use, e.g.\ \ttcode{-j4} for 4 cores.

\medskip
\begin{itemize}
    \item \ttcode{-DCMAKE\_BUILD\_TYPE=Release}. Only set to \ttcode{Debug} if you're a developer.
    \item \ttcode{-DCMAKE\_INSTALL\_PREFIX}. By default, executables install to \ttcode{\$\{CMAKE\_SOURCE\_DIR\}/\$\{CMAKE\_INSTALL\_BINDIR\}/}, headers to \ttcode{\$\{CMAKE\_INSTALL\_INCLUDEDIR\}/paratoric}, and static libraries to \ttcode{\$\{CMAKE\_SOURCE\_DIR\}/\$\{CMAKE\_INSTALL\_LIBDIR\}/}. The Python scripts expect \ttcode{\$\{CMAKE\_SOURCE\_DIR\}/bin/}; this directory always contains the \ttcode{paratoric} executable. To install into a custom directory, pass it via \ttcode{-DCMAKE\_INSTALL\_PREFIX}, e.g. \ttcode{-DCMAKE\_INSTALL\_PREFIX=/your/custom/directory/}.
    \item \ttcode{-DPARATORIC\_EXPORT\_COMPILE\_COMMANDS=ON}. Export \ttcode{compile\_commands.json} for tooling.
    \item \ttcode{-DPARATORIC\_LINK\_MPI=OFF}. Link the core to MPI, which is required on some clusters. The core itself does not need MPI.
    \item \ttcode{-DPARATORIC\_ENABLE\_NATIVE\_OPT=OFF}. Use \ttcode{-march=native} on GCC and Clang.
    \item \ttcode{-DPARATORIC\_ENABLE\_AVX2=OFF}. Use AVX2 (Haswell New Instructions). Requires a CPU which supports AVX2.
    \item \ttcode{-DPARATORIC\_ENABLE\_FAST\_MATH=ON}. Use \ttcode{-ffast-math} on GCC and Clang.
    \item \ttcode{-DPARATORIC\_BUILD\_TESTS=PROJECT\_IS\_TOP\_LEVEL}. Compile the tests (recommended).
    \item \ttcode{-DPARATORIC\_BUILD\_PYBIND=OFF}. Compile Python bindings.
    \item \ttcode{-DPARATORIC\_INSTALL\_TO\_SITE=OFF}. Install ParaToric Python module to site packages.
    \item \ttcode{-DPARATORIC\_PIP\_EDITABLE\_INSTALL=OFF}. Install ParaToric Python module via pip as an editable module.
    \item \ttcode{-DPARATORIC\_PIP\_OFFLINE\_INSTALL=OFF}. Turn on when installing to pip without internet access. Requires \ttcode{NumPy} and \ttcode{setuptools}.
\end{itemize}

\subsubsection{Module layout}
\begin{itemize}
  \item \ttcode{paratoric.\_paratoric}: compiled extension (PyBind11). Submodule: \ttcode{extended\_toric\_code}.
  \item \ttcode{paratoric.extended\_toric\_code}: convenient alias
  \item Running \ttcode{python \ -m \ paratoric} enters the package entry point (\ttcode{\_\_main\_\_.py}).
\end{itemize}

\subsubsection*{NumPy return formats}

All time series with potentially complex values are returned as \ttcode{complex128}. Real observables appear with zero imaginary part. Shapes are documented in the function references below.

\subsubsection*{API reference (\ttcode{paratoric.extended\_toric\_code})}

\paragraph{\ttcode{get\_thermalization(\dots{})}}
Run only the warmup and return per-snapshot observables and MC acceptance ratios.%
\ Internally converts \ttcode{std::variant<complex<double>,double>} to \ttcode{complex128} and \ttcode{std::vector<double>} to \ttcode{float64} arrays.%
\ The GIL is released while the C\texttt{++} routine executes.

\begin{center}
\small
\begin{tabularx}{\linewidth}{
  >{\raggedright\arraybackslash}p{0.35\linewidth}
  >{\raggedright\arraybackslash}p{0.20\linewidth}
  X
}
\hline
\textbf{Parameter} & \textbf{Type / default} & \textbf{Meaning}\\
\hline
\ttcode{N\_thermalization} & \ttcode{int} & Warmup steps.\\
\ttcode{N\_resamples} & \ttcode{int=1000} & Bootstrap resamples.\\
\ttcode{observables} & \ttcode{list[str]} & Names per snapshot.\\
\ttcode{seed} & \ttcode{int=0} & PRNG seed (\ttcode{0} $\Rightarrow$ random).\\
\ttcode{mu,h,J,lmbda} & \ttcode{float} & Hamiltonian parameters.\\
\ttcode{basis} & \ttcode{\{'x','z'\}='x'} & Spin eigenbasis.\\
\ttcode{lattice\_type} & \ttcode{str} & E.g.\ \ttcode{"triangular"}, \ttcode{"square"}, \dots\\
\ttcode{system\_size} & \ttcode{int} & Linear size per dimension.\\
\ttcode{beta} & \ttcode{float} & Inverse temperature.\\
\ttcode{boundaries} & \ttcode{str="periodic"} & Boundary condition.\\
\ttcode{default\_spin} & \ttcode{int=1} & Initial link spin (\ttcode{+1}/\ttcode{-1}).\\
\ttcode{save\_snapshots} & \ttcode{bool=false} & Enable snapshot files.\\
\ttcode{path\_out} & \ttcode{path|None=None} & Output directory (if saving).\\
\hline
\end{tabularx}
\end{center}

\textbf{Returns:} \ttcode{(series,\ acc\_ratio)}  
\quad \ttcode{series}: \ttcode{ndarray(complex128)} of shape \ttcode{(n\_obs,\ N\_thermalization)}; \ \ttcode{acc\_ratio}: \ttcode{ndarray(float64)} of shape \ttcode{(N\_thermalization,)}.

\medskip
\paragraph{\ttcode{get\_sample(\dots{})}}
Run thermalization and production sampling; return series and bootstrap statistics.%
\ Converts nested C\texttt{++} containers to NumPy arrays and releases the GIL during computation.

\begin{center}
\small
\begin{tabularx}{\linewidth}{
  >{\raggedright\arraybackslash}p{0.38\linewidth}
  >{\raggedright\arraybackslash}p{0.22\linewidth}
  X
}
\hline
\textbf{Parameter} & \textbf{Type / default} & \textbf{Meaning}\\
\hline
\ttcode{N\_samples} & \ttcode{int} & Stored samples per observable.\\
\ttcode{N\_thermalization} & \ttcode{int} & Warmup steps before sampling.\\
\ttcode{N\_between\_samples} & \ttcode{int} & Thinning between samples.\\
\ttcode{N\_resamples} & \ttcode{int=1000} & Bootstrap resamples.\\
\ttcode{custom\_therm} & \ttcode{bool=false} & Use \ttcode{h\_therm}, \ttcode{lmbda\_therm} during warmup.\\
\ttcode{observables} & \ttcode{list[str]} & Names per snapshot.\\
\ttcode{seed} & \ttcode{int=0} & PRNG seed (\ttcode{0} $\Rightarrow$ random).\\
\ttcode{mu,h,J,lmbda} & \ttcode{float} & Hamiltonian parameters.\\
\ttcode{h\_therm,lmbda\_therm} & \ttcode{float=0} & Warmup parameters if using custom thermalization.\\
\ttcode{basis} & \ttcode{\{'x','z'\}='x'} & Spin eigenbasis.\\
\ttcode{lattice\_type,system\_size,beta} & \ttcode{str,int,float} & Lattice and temperature.\\
\ttcode{boundaries,default\_spin} & \ttcode{str,int=("periodic",1)} & BC and initial spin.\\
\ttcode{save\_snapshots,path\_out} & \ttcode{bool=False,path|None=None} & Optional I/O.\\
\hline
\end{tabularx}
\end{center}

\textbf{Returns:} tuple of six arrays  
\quad \ttcode{series} \ttcode{(complex128)}: \ttcode{(n\_obs,\ N\_samples)}; \ 
\ttcode{mean}, \ttcode{mean\_std}, \ttcode{binder}, \ttcode{binder\_std}, \ttcode{tau\_int} \ttcode{(float64)}: each \ttcode{(n\_obs,)}.

\medskip
\paragraph{\ttcode{get\_hysteresis(\dots{})}}
Run a sweep where each step uses the previous state as its initial condition.%
\ Returns stacked arrays across steps; path handling validates per-step output directories when saving snapshots.

\begin{center}
\small
\begin{tabularx}{\linewidth}{
  >{\raggedright\arraybackslash}p{0.40\linewidth}
  >{\raggedright\arraybackslash}p{0.24\linewidth}
  X
}
\hline
\textbf{Parameter} & \textbf{Type / default} & \textbf{Meaning}\\
\hline
\ttcode{N\_samples,N\_thermalization,N\_between\_samples} & \ttcode{int,int,int} & Cadence per step.\\
\ttcode{N\_resamples} & \ttcode{int=1000} & Bootstrap resamples.\\
\ttcode{observables} & \ttcode{list[str]} & Names per snapshot.\\
\ttcode{seed} & \ttcode{int=0} & PRNG seed.\\
\ttcode{mu,J} & \ttcode{float} & Star and plaquette couplings.\\
\ttcode{h\_hys,lmbda\_hys} & \ttcode{list[float]} & Hysteresis values, length must match.\\
\ttcode{basis} & \ttcode{\{'x','z'\}='x'} & Spin basis.\\
\ttcode{lattice\_type,system\_size,beta} & \ttcode{str,int,float} & Lattice and temperature.\\
\ttcode{boundaries,default\_spin} & \ttcode{str,int=("periodic",1)} & BC and initial spin.\\
\ttcode{save\_snapshots} & \ttcode{bool=false} & Enable stepwise I/O.\\
\ttcode{paths\_out} & \ttcode{list[path]|None=None} & Output path per step (size must match \ttcode{h\_hys} if saving).\\
\hline
\end{tabularx}
\end{center}

\textbf{Returns:} tuple of six arrays  
\quad \ttcode{series3d} \ttcode{(complex128)}: \ttcode{(n\_steps,\ n\_obs,\ N\_samples)}; \
\ttcode{mean2d}, \ttcode{std2d}, \ttcode{binder2d}, \ttcode{binder\_std2d}, \ttcode{tau2d} \ttcode{(float64)}: each \ttcode{(n\_steps,\ n\_obs)}.%
\ The number of steps equals \ttcode{len(h\_hys)} (and \ttcode{len(lmbda\_hys)}).

\subsubsection*{Array dtypes and shapes (summary)}
\begin{center}
\small
\begin{tabularx}{\linewidth}{
  >{\raggedright\arraybackslash}p{0.26\linewidth}
  >{\raggedright\arraybackslash}p{0.32\linewidth}
  X
}
\hline
\textbf{Function} & \textbf{Name / dtype} & \textbf{Shape}\\
\hline
\ttcode{get\_thermalization} & \ttcode{series} (\ttcode{complex128}) & \ttcode{(n\_obs,\ N\_thermalization)}\\
 & \ttcode{acc\_ratio} (\ttcode{float64}) & \ttcode{(N\_thermalization,)}\\
\ttcode{get\_sample} & \ttcode{series} (\ttcode{complex128}) & \ttcode{(n\_obs,\ N\_samples)}\\
 & \mbox{\ttcode{mean}}, \mbox{\ttcode{mean\_std}}, \mbox{\ttcode{binder}}, \mbox{\ttcode{binder\_std}}, \ttcode{tau\_int} \mbox{(\ttcode{float64})} & each \ttcode{(n\_obs,)}\\
\ttcode{get\_hysteresis} & \ttcode{series3d} (\ttcode{complex128}) & \ttcode{(n\_steps,\ n\_obs,\ N\_samples)}\\
 & \mbox{\ttcode{mean2d}}, \mbox{\ttcode{std2d}}, \mbox{\ttcode{binder2d}}, \mbox{\ttcode{binder\_std2d}}, \mbox{\ttcode{tau2d}} \mbox{(\ttcode{float64})} & each \ttcode{(n\_steps,\ n\_obs)}\\
\hline
\end{tabularx}
\end{center}

\subsubsection*{Notes on performance}

The bindings release the global interpreter lock (GIL) during heavy compute (\ttcode{py::gil\_scoped\_release}), enabling multi-threaded C\texttt{++} execution if the backend uses threads or when calling from multiprocessing workers. Conversions handle 1D/2D/3D containers and enforce consistent inner lengths before copying to NumPy.

\subsubsection{Usage example}

\begin{lstlisting}[language=python, basicstyle=\ttfamily\small\color{darkblue}, caption={Importing and calling from Python}]
>>> import numpy as np
>>> from paratoric import extended_toric_code as etc
>>> series, acc_ratio = etc.get_thermalization(
...     N_thermalization=2000, N_resamples=500,
...     observables=["energy","plaquette_z","anyon_count"],
...     seed=0, mu=1.0, h=0.2, J=1.0, lmbda=0.0,
...     basis='z', lattice_type="square", system_size=16, beta=8.0,
...     boundaries="periodic", default_spin=1,
...     save_snapshots=False, path_out=None)
>>> series.shape, series.dtype
((3, 2000), dtype('complex128'))
>>> out = etc.get_sample(
...     N_samples=1000, N_thermalization=5000, N_between_samples=10,
...     N_resamples=1000, custom_therm=False,
...     observables=["energy","plaquette_z"],
...     seed=0, mu=1.0, h=0.2, h_therm=0.0,
...     J=1.0, lmbda=0.0, lmbda_therm=0.0,
...     basis='z', lattice_type="square", system_size=16, beta=8.0,
...     boundaries="periodic", default_spin=1,
...     save_snapshots=False, path_out=None)
>>> (series_s, mean, mean_std, binder, binder_std, tau_int) = out
\end{lstlisting}

\subsection{Python command-line interface}

ParaToric ships a Python command-line interface \texttt{/python/cli/paratoric.py} that orchestrates C\texttt{++} backends, runs sweeps, writes HDF5/XML outputs and plots observables, Binder ratios, and integrated autocorrelation times.  
It requires \href{https://numpy.org/}{NumPy} \cite{Harris2020}, \href{https://matplotlib.org/}{Matplotlib} \cite{Hunter2007}, and \href{https://www.h5py.org/}{H5py}.

\subsubsection{Build \& Installation}

The command-line interface requires \href{https://www.hdfgroup.org/}{HDF5} $\geq$1.14.3 (older HDF5 versions may work, but were not tested). The core requires C++23, \href{https://cmake.org/}{CMake} $\geq$3.23 and \href{https://www.boost.org/}{Boost} $\geq$1.87 (older Boost versions may work, but were not tested). To compile it, run:

\begin{verbatim}
cmake -S . -B build -DCMAKE_BUILD_TYPE=Release \
-DPARATORIC_ENABLE_NATIVE_OPT=ON -DPARATORIC_LINK_MPI=OFF \
-DPARATORIC_BUILD_TESTS=ON -DPARATORIC_BUILD_CLI=ON
cmake --build build -jN
ctest --test-dir build -jN --output-on-failure
cmake --install build
\end{verbatim}

Replace \ttcode{N} with the number of cores to use, e.g.\ \ttcode{-j4} for 4 cores.

\medskip
\begin{itemize}
    \item \ttcode{-DCMAKE\_BUILD\_TYPE=Release}. Only set to \ttcode{Debug} if you're a developer.
    \item \ttcode{-DCMAKE\_INSTALL\_PREFIX}. By default, executables install to \ttcode{\$\{CMAKE\_SOURCE\_DIR\}/\$\{CMAKE\_INSTALL\_BINDIR\}/}, headers to \ttcode{\$\{CMAKE\_INSTALL\_INCLUDEDIR\}/paratoric}, and static libraries to \ttcode{\$\{CMAKE\_SOURCE\_DIR\}/\$\{CMAKE\_INSTALL\_LIBDIR\}/}. The Python scripts expect \ttcode{\$\{CMAKE\_SOURCE\_DIR\}/bin/}; this directory always contains the \ttcode{paratoric} executable. To install into a custom directory, pass it via \ttcode{-DCMAKE\_INSTALL\_PREFIX}, e.g. \ttcode{-DCMAKE\_INSTALL\_PREFIX=/your/custom/directory/}.
    \item \ttcode{-DPARATORIC\_EXPORT\_COMPILE\_COMMANDS=ON}. Export \ttcode{compile\_commands.json} for tooling.
    \item \ttcode{-DPARATORIC\_LINK\_MPI=OFF}. Link the core to MPI, which is required on some clusters. The core itself does not need MPI.
    \item \ttcode{-DPARATORIC\_ENABLE\_NATIVE\_OPT=OFF}. Use \ttcode{-march=native} on GCC and Clang.
    \item \ttcode{-DPARATORIC\_ENABLE\_AVX2=OFF}. Use AVX2 (Haswell New Instructions). Requires a CPU which supports AVX2.
    \item \ttcode{-DPARATORIC\_ENABLE\_FAST\_MATH=ON}. Use \ttcode{-ffast-math} on GCC and Clang.
    \item \ttcode{-DPARATORIC\_BUILD\_TESTS=PROJECT\_IS\_TOP\_LEVEL}. Compile the tests (recommended).
    \item \ttcode{-DPARATORIC\_BUILD\_CLI=PROJECT\_IS\_TOP\_LEVEL} Required for both the C++ and Python command line interface.
\end{itemize}

\subsubsection*{General options}
\begin{center}
\small
\begin{tabularx}{\linewidth}{
  >{\raggedright\arraybackslash}p{0.32\linewidth}
  >{\raggedright\arraybackslash}p{0.14\linewidth}
  X
}
\hline
\textbf{Long flag} & \textbf{Short} & \textbf{Description}\\
\hline
\ttcode{--help} & \ttcode{-h} & Show help and exit.\\
\ttcode{--simulation} & \ttcode{-sim} & Simulation type selector.\\
\ttcode{--N\_thermalization} & \ttcode{-Nth} & Thermalization steps (proposed updates).\\
\ttcode{--N\_samples} & \ttcode{-Ns} & Number of samples/snapshots.\\
\ttcode{--N\_between\_steps} & \ttcode{-Nbs} & Steps between successive samples (thinning).\\
\ttcode{--N\_resamples} & \ttcode{-Nr} & Bootstrap resamples.\\
\ttcode{--custom\_therm} & \ttcode{-cth} & Use thermalization values for $h,\lambda$ (\ttcode{0} or \ttcode{1}).\\
\ttcode{--observables} & \ttcode{-obs} & Space-separated list, e.g.\ \ttcode{fredenhagen\_marcu \ percolation\_probability \ energy}.\\
\ttcode{--seed} & \ttcode{-seed} & PRNG seed; \ttcode{0} means random seed.\\
\ttcode{--mu\_constant} & \ttcode{-muc} & Value of $\mu$.\\
\ttcode{--J\_constant} & \ttcode{-Jc} & Value of $J$.\\
\ttcode{--h\_constant} & \ttcode{-hc} & Value of $h$.\\
\ttcode{--h\_constant\_therm} & \ttcode{-hct} & Thermalization value of $h$.\\
\ttcode{--lmbda\_constant} & \ttcode{-lmbdac} & Value of $\lambda$.\\
\ttcode{--lmbda\_constant\_therm} & \ttcode{-lmbdact} & Thermalization value of $\lambda$.\\
\ttcode{--output\_directory} & \ttcode{-outdir} & Output directory.\\
\ttcode{--snapshots} & \ttcode{-snap} & Save snapshots toggle (\ttcode{0}/\ttcode{1}).\\
\ttcode{--full\_time\_series} & \ttcode{-fts} & Save full time series toggle (\ttcode{0}/\ttcode{1}).\\
\ttcode{--processes} & \ttcode{-proc} & Logical CPU count for Python multiprocessing. 0 means all available cores. Negative numbers $-x$ mean use all cores minus x. Default is -4.\\
\hline
\end{tabularx}
\end{center}

\subsubsection*{Lattice-specific options}
\begin{center}
\small
\begin{tabularx}{\linewidth}{
  >{\raggedright\arraybackslash}p{0.32\linewidth}
  >{\raggedright\arraybackslash}p{0.14\linewidth}
  X
}
\hline
\textbf{Long flag} & \textbf{Short} & \textbf{Description}\\
\hline
\ttcode{--help} & \ttcode{-h} & Show help and exit.\\
\ttcode{--basis} & \ttcode{-bas} & Spin basis: \ttcode{x} or \ttcode{z}.\\
\ttcode{--lattice\_type} & \ttcode{-lat} & \ttcode{square}, \ttcode{cubic}, \ttcode{triangular}, \ttcode{honeycomb}, \dots\\
\ttcode{--system\_size} & \ttcode{-L} & Linear size; in 2D, \ttcode{30} yields a $30\times 30$ lattice (unit cells).\\
\ttcode{--temperature} & \ttcode{-T} & Temperature $T=1/\beta>0$.\\
\ttcode{--boundaries} & \ttcode{-bound} & \ttcode{periodic} or \ttcode{open}.\\
\ttcode{--default\_spin} & \ttcode{-dsp} & Initial edge spin: \ttcode{1} or \ttcode{-1}.\\
\hline
\end{tabularx}
\end{center}

\bigskip
The command line interface offers several sweep modes. All are embarrassingly parallel; set \ttcode{--processes} close to the number of steps when possible.

\subsubsection{\texorpdfstring{$T$-sweep}{T-sweep}}
Runs \ttcode{T\_steps} independent Markov chains for evenly spaced temperatures $T \in$ [\ttcode{T\_lower}, \\ \ttcode{T\_upper}].

\begin{lstlisting}[language=bash, basicstyle=\ttfamily\small\color{darkblue}, breaklines=true, caption={Example usage}]
python3 ./python/cli/paratoric.py -sim etc_T_sweep -Nbs 8000 -Ns 1000 -muc 1 -Nth 50000 -Tl 0.1 -Tu 10 -Ts 15 -hc 0.5 -Jc 1 -lmbdac 0.2 -Nr 1000 -obs percolation_strength percolation_probability largest_cluster largest_plaquette_cluster string_number energy energy_h energy_mu energy_J energy_lmbda sigma_x sigma_z star_x plaquette_z staggered_imaginary_times delta anyon_count anyon_density fredenhagen_marcu sigma_x_static_susceptibility sigma_x_dynamical_susceptibility sigma_z_static_susceptibility sigma_z_dynamical_susceptibility -bas x -lat square -L 6 -bound periodic -dsp 1 -outdir /your/output/directory 
\end{lstlisting}

\paragraph{Sweep-specific flags.}
\begin{center}
\small
\begin{tabularx}{\linewidth}{
  >{\raggedright\arraybackslash}p{0.32\linewidth}
  >{\raggedright\arraybackslash}p{0.14\linewidth}
  X
}
\hline
\textbf{Long flag} & \textbf{Short} & \textbf{Description}\\
\hline
\ttcode{--simulation} & \ttcode{-sim} & Use \ttcode{etc\_T\_sweep}.\\
\ttcode{--T\_lower} & \ttcode{-Tl} & Lower bound of $T$.\\
\ttcode{--T\_upper} & \ttcode{-Tu} & Upper bound of $T$.\\
\ttcode{--T\_steps} & \ttcode{-Ts} & Number of temperatures between bounds.\\
\hline
\end{tabularx}
\end{center}

\subsubsection{h-sweep}
Runs \ttcode{h\_steps} independent chains in parallel for evenly spaced $h \in$ [\ttcode{h\_lower}, \ttcode{h\_upper}].

\begin{lstlisting}[language=bash, basicstyle=\ttfamily\small\color{darkblue}, breaklines=true, caption={Example usage}]
python3 ./python/cli/paratoric.py -sim etc_h_sweep -Nbs 8000 -Ns 1000 -muc 1 -Nth 50000 -hl 0.0 -hu 0.5 -hs 15 -T 0.1 -Jc 1 -lmbdac 0.2 -Nr 1000 -obs percolation_strength percolation_probability plaquette_percolation_probability plaquette_percolation_strength largest_cluster largest_plaquette_cluster string_number energy energy_h energy_mu energy_J energy_lmbda sigma_x sigma_z star_x plaquette_z staggered_imaginary_times delta anyon_count anyon_density fredenhagen_marcu sigma_x_static_susceptibility sigma_x_dynamical_susceptibility sigma_z_static_susceptibility sigma_z_dynamical_susceptibility -bas x -lat square -L 6 -bound periodic -dsp 1 -outdir /your/output/directory
\end{lstlisting}

\paragraph{Sweep-specific flags.}
\begin{center}
\small
\begin{tabularx}{\linewidth}{
  >{\raggedright\arraybackslash}p{0.32\linewidth}
  >{\raggedright\arraybackslash}p{0.14\linewidth}
  X
}
\hline
\textbf{Long flag} & \textbf{Short} & \textbf{Description}\\
\hline
\ttcode{--simulation} & \ttcode{-sim} & Use \ttcode{etc\_h\_sweep}.\\
\ttcode{--h\_lower} & \ttcode{-hl} & Lower bound of $h$.\\
\ttcode{--h\_upper} & \ttcode{-hu} & Upper bound of $h$.\\
\ttcode{--h\_steps} & \ttcode{-hs} & Number of field steps between bounds.\\
\hline
\end{tabularx}
\end{center}

\subsubsection{\texorpdfstring{$\lambda$-sweep}{lambda-sweep}}
Runs \ttcode{lmbda\_steps} independent chains in parallel for evenly spaced $\lambda\in$ [\ttcode{lmbda\_lower}, \\ \ttcode{lmbda\_upper}].

\begin{lstlisting}[language=bash, basicstyle=\ttfamily\small\color{darkblue}, breaklines=true, caption={Example usage}]
python3 ./python/cli/paratoric.py -sim etc_lmbda_sweep -Nbs 12000 -Ns 1000 -muc 1 -Nth 100000 -lmbdal 0.1 -lmbdau 0.7 -lmbdas 15 -T 0.1 -hc 0.1 -Jc 1 -Nr 1000 -obs percolation_strength percolation_probability plaquette_percolation_probability plaquette_percolation_strength largest_cluster largest_plaquette_cluster string_number energy energy_h energy_mu energy_J energy_lmbda sigma_x sigma_z star_x plaquette_z staggered_imaginary_times delta anyon_count anyon_density fredenhagen_marcu sigma_x_static_susceptibility sigma_x_dynamical_susceptibility sigma_z_static_susceptibility sigma_z_dynamical_susceptibility -bas z -lat honeycomb -L 6 -bound periodic -dsp 1 -outdir /your/output/directory 
\end{lstlisting}

\paragraph{Sweep-specific flags.}
\begin{center}
\small
\begin{tabularx}{\linewidth}{
  >{\raggedright\arraybackslash}p{0.32\linewidth}
  >{\raggedright\arraybackslash}p{0.14\linewidth}
  X
}
\hline
\textbf{Long flag} & \textbf{Short} & \textbf{Description}\\
\hline
\ttcode{--simulation} & \ttcode{-sim} & Use \ttcode{etc\_lmbda\_sweep}.\\
\ttcode{--lmbda\_lower} & \ttcode{-lmbdal} & Lower bound of $\lambda$.\\
\ttcode{--lmbda\_upper} & \ttcode{-lmbdau} & Upper bound of $\lambda$.\\
\ttcode{--lmbda\_steps} & \ttcode{-lmbdas} & Number of field steps between bounds.\\
\hline
\end{tabularx}
\end{center}

\subsubsection{\texorpdfstring{circle-sweep}{circle-sweep}}
Runs \ttcode{Theta\_steps} independent chains in parallel along a circle in $(\lambda,h)$ centered at \\ \ttcode{(lmbda\_constant, h\_constant)} with radius \ttcode{radius}, for angles $\Theta\in[$\ttcode{Theta\_lower},\\ \ttcode{Theta\_upper}$]$ (angles measured anti-clockwise from the $\lambda$-axis, i.e., \\ $\lambda=\,$\ttcode{lmbda\_constant}+\ttcode{radius} $\cos\Theta$, $h=\,$\ttcode{h\_constant}+\ttcode{radius} $\sin\Theta$).

\begin{lstlisting}[language=bash, basicstyle=\ttfamily\small\color{darkblue}, breaklines=true, caption={Example usage}]
python3 ./python/cli/paratoric.py -sim etc_circle_sweep -Nbs 8000 -Ns 1000 -muc 1 -Nth 50000 -lmbdac 0.0 -rad 1.0 -Thl 0 -Thu 1.57 -Ths 30 -T 0.1 -hc 0.0 -Jc 1 -Nr 1000 -obs percolation_strength percolation_probability largest_cluster largest_plaquette_cluster string_number energy energy_h energy_mu energy_J energy_lmbda sigma_x sigma_z star_x plaquette_z staggered_imaginary_times delta anyon_count anyon_density fredenhagen_marcu sigma_x_static_susceptibility sigma_x_dynamical_susceptibility sigma_z_static_susceptibility sigma_z_dynamical_susceptibility -bas x -lat square -L 6 -bound periodic -dsp 1 -outdir /your/output/directory 
\end{lstlisting}

\paragraph{Sweep-specific flags.}
\begin{center}
\small
\begin{tabularx}{\linewidth}{
  >{\raggedright\arraybackslash}p{0.32\linewidth}
  >{\raggedright\arraybackslash}p{0.14\linewidth}
  X
}
\hline
\textbf{Long flag} & \textbf{Short} & \textbf{Description}\\
\hline
\ttcode{--simulation} & \ttcode{-sim} & Use \ttcode{etc\_circle\_sweep}.\\
\ttcode{--lmbda\_constant} & \ttcode{-lmbdac} & Circle center in $\lambda$.\\
\ttcode{--h\_constant} & \ttcode{-hc} & Circle center in $h$.\\
\ttcode{--radius} & \ttcode{-rad} & Circle radius.\\
\ttcode{--Theta\_lower} & \ttcode{-Thl} & Lower bound of $\Theta$.\\
\ttcode{--Theta\_upper} & \ttcode{-Thu} & Upper bound of $\Theta$.\\
\ttcode{--Theta\_steps} & \ttcode{-Ths} & Number of angles between bounds.\\
\hline
\end{tabularx}
\end{center}


\subsubsection{\texorpdfstring{Hysteresis-sweep}{Hysteresis-sweep}}
Uses the hysteresis schedule specified in \ttcode{hhys} and \ttcode{lmbdahys}. This mode will run two Markov chains, one in the original parameter order specified in \ttcode{hhys} and \ttcode{lmbdahys}, and one with a reversed parameter order, i.e., it calculates both branches of the hysteresis loop.

\begin{lstlisting}[language=bash, basicstyle=\ttfamily\small\color{darkblue}, breaklines=true, caption={Example usage}]
python3 ./python/cli/paratoric.py -sim etc_hysteresis -Nbs 5000 -Ns 1000 -muc 1 -Nth 50000 -hhys 0.5 0.6 0.7 0.8 0.9 1.0 1.1 1.2 1.3 1.4 1.5 -T 0.1 -Jc 1 -lmbdahys 0 0 0 0 0 0 0 0 0 0 0 -Nr 1000 -obs percolation_strength percolation_probability largest_cluster largest_plaquette_cluster string_number energy energy_h energy_mu energy_J energy_lmbda sigma_x sigma_z star_x plaquette_z staggered_imaginary_times delta anyon_count anyon_density fredenhagen_marcu sigma_x_static_susceptibility sigma_x_dynamical_susceptibility sigma_z_static_susceptibility sigma_z_dynamical_susceptibility -bas x -lat cubic -L 4 -bound periodic -dsp 1 -outdir /your/output/directory 
\end{lstlisting}

\paragraph{Sweep-specific flags.}
\begin{center}
\small
\begin{tabularx}{\linewidth}{
  >{\raggedright\arraybackslash}p{0.32\linewidth}
  >{\raggedright\arraybackslash}p{0.14\linewidth}
  X
}
\hline
\textbf{Long flag} & \textbf{Short} & \textbf{Description}\\
\hline
\ttcode{--simulation} & \ttcode{-sim} & Use \ttcode{etc\_hysteresis}.\\
\ttcode{--lmbda\_hysteresis} & \ttcode{-lmbdahys} & Hysteresis schedule for $\lambda$. Length must match \mbox{\ttcode{hhys}}.\\
\ttcode{--h\_hysteresis} & \ttcode{-hhys} & Hysteresis schedule for $h$. Length must match \mbox{\ttcode{lmbdahys}}.\\
\hline
\end{tabularx}
\end{center}

\subsubsection{Thermalization}
Runs \ttcode{repetitions} independent chains in parallel and reports observables and MC acceptance ratios every step, averaged over chains.

\begin{lstlisting}[language=bash, basicstyle=\ttfamily\small\color{darkblue}, breaklines=true, caption={Example usage}]
python3 ./python/cli/paratoric.py -sim etc_thermalization -muc 1 -Nth 50000 -reps 10 -lmbdac 2 -T 0.1 -hc 0.3 -Jc 1 -Nr 1000 -obs percolation_strength percolation_probability plaquette_percolation_strength plaquette_percolation_probability largest_cluster largest_plaquette_cluster string_number energy energy_h energy_mu energy_J energy_lmbda sigma_x sigma_z star_x plaquette_z staggered_imaginary_times delta anyon_count anyon_density fredenhagen_marcu sigma_x_static_susceptibility sigma_x_dynamical_susceptibility sigma_z_static_susceptibility sigma_z_dynamical_susceptibility -bas x -lat square -L 4 -bound periodic -dsp 1 -outdir /your/output/directory 
\end{lstlisting}

\paragraph{Sweep-specific flags.}
\begin{center}
\small
\begin{tabularx}{\linewidth}{
  >{\raggedright\arraybackslash}p{0.32\linewidth}
  >{\raggedright\arraybackslash}p{0.14\linewidth}
  X
}
\hline
\textbf{Long flag} & \textbf{Short} & \textbf{Description}\\
\hline
\ttcode{--simulation} & \ttcode{-sim} & Use \ttcode{etc\_thermalization}.\\
\ttcode{--repetitions} & \ttcode{-reps} & Number of Markov chains to average.\\
\hline
\end{tabularx}
\end{center}

\section{Using ParaToric}

\subsection{Monte Carlo Updates}

There is no need for the user to explicitly call specific updates or interact with internal C++ classes when using the documented interfaces. Internally, we use all five updates described in the original algorithm of Wu, Deng, and Prokof'ev \cite{Wu2012}. These must furthermore be supplemented with the following two updates: Because for high temperatures and for zero off-diagonal fields the spin at imaginary time $0=\beta$ cannot be flipped, we allow for flipping the spin on the entire imaginary axis on one bond or on a plaquette (star) in the $\hat{\sigma}^x$-basis ($\hat{\sigma}^z$-basis).

These updates only change the energy terms diagonal in the given basis and are trivial when caching the total integrated diagonal energy (the update locally flips the sign of the total integrated potential energy). Another advantage is that integrated autocorrelation times for observables diagonal in the given basis improve even in regimes that were previously accessible. By default, all seven updates are equally likely to be proposed. The user may manually fine-tune the proposal probabilities when maximizing performance in a very specific parameter regime. We use a 64-bit Mersenne-Twister for pseudorandom numbers \cite{Matsumoto1998} with the ability to set the seed externally. Some updates have early exits for input parameters for which they will always be rejected.

\subsection{Monte Carlo Diagnostics} \label{sec:diagnostics}

There are two compilation modes, \ttcode{Release} and \ttcode{Debug}. In production runs, one should always use the \ttcode{Release} mode; however, it still gives the user enough information to diagnose sampling problems without severe performance impacts.

\subsubsection{Thermalization mode}
We provide thermalization routines which should be used before production runs to ensure proper thermalization (also known as burn-in). Thermalization times can vary drastically between different observables and initial conditions. We provide an example of sufficient and insufficient thermalization in Fig.~\ref{fig:Thermalization}. We recommend using the provided Python command-line interface, which will also plot the thermalization of all measured observables for the user.

In thermalization runs, we also return the Monte Carlo acceptance ratio of every update. This can also be used to diagnose freezing (in the measurement phase, use the integrated autocorrelation time instead), e.g., when the acceptance ratio is always identical and/or very low. 

In case one suspects experiencing a serious sampling problem, we recommend recompiling the project in the \ttcode{Debug} mode, which provides a wide array of runtime debug information about the proposed steps, acceptance ratios, and intermediate results. However, do not use the \ttcode{Debug} mode in production runs, as it negatively impacts performance.

\subsubsection{Integrated autocorrelation time} 
When measuring observables, we first thermalize the system with \ttcode{N\_thermalization} steps, then measure \ttcode{N\_samples} times with \ttcode{N\_between\_samples} steps between measurements. The normalized autocorrelation function $\rho_O(k)$ of an observable $O_k$ (observable $O$ measured at time $k$) applied to a discrete time series of length $N$ is given by:
\begin{align}
    \rho_O(k) = \frac{C(k)}{C(0)}, \qquad C(k) = \frac{1}{N-k} \sum_{i=0}^{N-k-1} (O_i-\bar{O})(O_{i+k}-\bar{O}), \qquad \bar{O}=\frac{1}{N} \sum_{i=0}^{N-1} O_i.
\end{align}
It is a statistical measure of the correlations between measurements of observable $O$ at times $i$ and $i+k$.\footnote{In ParaToric, the autocorrelation function is calculated efficiently using fast Fourier transforms.} We define the \textit{integrated autocorrelation time}
\begin{align}
    \tau^O_\mathrm{int} = \frac{1}{2} + \sum_{k\geq 1} \rho_O(k).
\end{align}
Large $\tau_\mathrm{int}$ are generally undesirable since they increase error bars and can lead to bias. 
In case of perfect sampling, we would have $\rho_O(0)=1$ and $\rho_O(k)=0 \; \forall k \geq 1$, i.e., each measurement is only correlated with itself but not with other measurements and $\tau_\mathrm{int} = 1/2$. In practice, this is usually not feasible, and we have to work with a finite autocorrelation time $\tau_\mathrm{int} > 1/2$. 

When using ParaToric, we strongly recommend monitoring $\tau_\mathrm{int}$ for all simulations and all observables. It is automatically calculated for every observable based on the full time series. As a rule of thumb, the autocorrelation is fine as long as $\tau_\mathrm{int} \ll N_\mathrm{samples}$, otherwise it leads to bias and seriously underestimated error bars. In the vicinity of phase transitions, $\tau_\mathrm{int}$ dramatically increases (``critical slowing down'') \cite{Wolff1990}. Importantly, $\tau_\mathrm{int}$ can differ vastly between different observables! If the autocorrelation is too high, increase the number of steps between samples. In more complicated cases, one may need to adapt the update proposal distributions and/or the updates themselves as a last resort.

It is also important to mention that ParaToric only computes a statistical \textit{estimate} of $\tau_\mathrm{int}$. Many factors determine how accurate this estimate is, and crucially, the system needs to be properly \textit{thermalized}. In principle, one can use $\tau_\mathrm{int}$ in the way it is computed above directly for calculating error bars of correlated time series; however, ParaToric uses a more robust bootstrapping approach.

\subsubsection{Error bars} 

ParaToric applies the stationary bootstrap \cite{Politis1994, Politis2004, Patton2009} for all error bars, thus capturing autocorrelation effects. Large $\tau_\mathrm{int}$ will lead to worse error bars. The only parameter that the user can change is the number of bootstrap resamples \ttcode{N\_resamples}. The default is 1000, which is enough in most cases. Note that a too low value of \ttcode{N\_between\_steps} increases the relative computational cost of performing the measurements, which may negatively affect the code efficiency at no statistical gain. If the error bars are too large, either the number of samples is too low (in which case one should increase \ttcode{N\_samples}) or the autocorrelation is too large (in which case one could additionally increase \ttcode{N\_between\_samples}).

\subsection{Tips \& tricks}

\subsubsection{Probing ground state physics} 
The algorithm implemented by ParaToric fundamentally requires a finite temperature $T>0$. However, in QMC simulations, there is always a finite-size energy gap (the difference between the energy of the ground state and the first excited state). Additionally, some phases like the topological ground state of the toric code have a physical bulk gap (even at $L \to \infty$). As long as the temperature is well below the total gap, we are exponentially close to the ground state. Usually, a temperature $T\sim1/L$ suffices for the toric code, although other situations may arise.

\subsubsection{Probing first-order transitions} 
ParaToric provides functionalities to probe weak and strong first-order phase transitions. The hysteresis mode can be used to probe hysteresis loops in the vicinity of strong first-order phase transitions, by repeating the simulation two times and mirroring the order of the parameters in \ttcode{h\_hysteresis} and \ttcode{lmbda\_hysteresis}. Weak first-order transitions can be detected by plotting a time series histogram of an observable (it exhibits a double-peak structure). Both approaches have been used in the context of the toric code \cite{Linsel2026}.

\subsubsection{Choosing the basis} Sometimes, one can work in both the $\hat{\sigma}^x$- and the $\hat{\sigma}^z$-basis. However, the performance can vary drastically! Generally, the $\hat{\sigma}^x$-basis is more efficient for $h/J > \lambda / \mu$ and vice versa. 

\subsubsection{Choosing \ttcode{N\_thermalization}} 
Based on our experience, we can say that $N_\mathrm{thermalization} = 500 L^d / T$ is a sensible choice for small fields, where $d$ is the dimensionality of the system.
Nevertheless, one should make use of the provided tools to benchmark thermalization, see Sec.~\ref{sec:diagnostics}, and rather err on the side of safety.

\subsubsection{Choosing \ttcode{N\_samples}} 
Neglecting autocorrelation effects, the error of an observable $\Delta O$ scales as \linebreak $\Delta O \sim 1/\sqrt{N_\mathrm{samples}}$. More samples are, in principle, always better and lead to lower error bars. Smoothness of a curve of statistical results also requires that error bars be small in relation to the parameter grid size. If one increases the parameter resolution (e.g., in the field $h$), then one typically also increases \ttcode{N\_samples}.

\subsubsection{Choosing \ttcode{N\_between\_samples}} 
The optimal choice for \ttcode{N\_samples} is the integrated autocorrelation time of the time series (where a measurement is taken after every update). A good guess of the autocorrelation time based on previous simulations for smaller system sizes or nearby parameter points can result in substantial computational saving costs in production runs for large system sizes. Near continuous phase transitions, the integrated autocorrelation time has an additional dependence $\tau_\mathrm{int} \sim L^z$, where $z$ is the dynamical exponent of the universality class of the transition. For a 2D system in the vicinity of a continuous phase transition, a sensible scaling for \ttcode{N\_between\_samples} could be $\mathcal{O}(L^2 \times \beta \times L^z)$ ($\mathcal{O}(L^2)$ links, each has off-diagonal spin flips $\mathcal{O}(\beta)$).

\subsubsection{Choosing \ttcode{N\_resamples}} 
As with \ttcode{N\_samples}, more is better (but also more costly). Usually \ttcode{N\_resamples} $\approx 1000$ is a sensible choice.

\subsubsection{Extracting snapshots} 
When the option \ttcode{save\_snapshots} is enabled, ParaToric will write the snapshots into the directory specified in \ttcode{path\_out} (or in the paths \ttcode{paths\_out} for hysteresis sweeps). The snapshots are saved in the GraphML format (XML-based), which is supported by many major graph libraries. One snapshot will be saved for every measurement of observables, i.e., \ttcode{N\_samples} snapshots in total. All snapshots are written into a single file to save disk space and simultaneously offer a structured, self-documenting format. Every edge stores a list of spins. The first spin belongs to the first snapshot, the second one to the second snapshot, and so on. There are no special requirements for disks or memory bandwidth; the snapshots are kept in RAM and are only written to disk after the simulation has finished.

\subsubsection{Adding new observables/lattices/updates} 
After adding features to the code, \textit{always} benchmark them using analytical results, other numerical methods (exact diagonalization, tensor networks, ...), and unit tests. We advise using a fixed seed during development, e.g., when checking whether two methods produce the exact same result. The code has some built-in features to check self-consistency, e.g., at the end of each simulation, the code checks whether the cached total energy is numerically close to the total energy calculated from scratch. Do not turn off these features, as they will point you toward bugs!

\subsection{Benchmarks} 

\subsubsection{Thermalization}

\begin{figure}[t]
\centering
\includegraphics[width=0.95\textwidth]{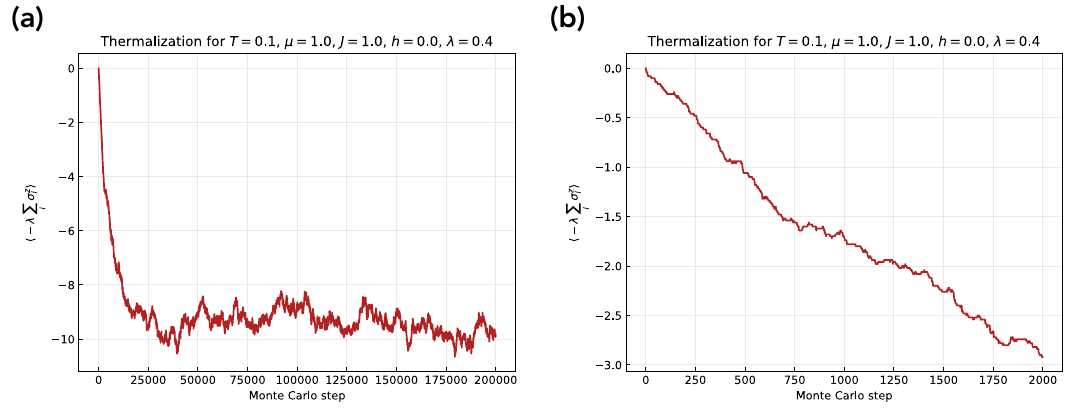}
\caption{\textbf{Good and bad thermalization} plots produced by the Python command-line interface. We show the gauge field energy $\propto \lambda$. (a) The system is well thermalized; after its initial drop, the energy fluctuates around the expectation value. (b) The system is not yet thermalized; the floating average is still decreasing. }
\label{fig:Thermalization}
\end{figure}

In Fig.~\ref{fig:Thermalization} (which we already discussed before and repeat here for completeness) we plot the gauge field energy $\propto \lambda$ for two systems: the left one is sufficiently thermalized, the right one is not. The plots are direct outputs of the Python command-line interface. Always make sure the system is well thermalized.

\subsubsection{Integrated autocorrelation time}

Here, we demonstrate how the integrated autocorrelation time $\tau_\mathrm{int}$ grows with decreasing \ttcode{N\_between\_samples}. We use the following setup:

\begin{lstlisting}[language=python, basicstyle=\ttfamily\small\color{darkblue}, caption={\ttcode{N\_between\_samples} benchmarking setup}]
>>> import numpy as np
>>> from paratoric import extended_toric_code as etc
>>> series, mean, std, binder, binder_std, tau_int = etc.get_sample(N_samples=100000, N_thermalization=10000, N_between_samples=1, N_resamples=1000, custom_therm=False, observables=["energy"], seed=0, mu=1.0, h=0.0, h_therm=0.0, J=1.0, lmbda=0, lmbda_therm=0.0, basis='x', lattice_type="square", system_size=4, beta=10, boundaries="periodic", default_spin=1, save_snapshots=False)
\end{lstlisting}

We only run the simulation once per \ttcode{N\_between\_samples}. The results are:

\begin{center}
\small
\begin{tabularx}{\linewidth}{
  >{\raggedright\arraybackslash}p{0.24\linewidth}
  >{\centering\arraybackslash}p{0.14\linewidth}
  >{\centering\arraybackslash}p{0.14\linewidth}
  >{\centering\arraybackslash}p{0.14\linewidth}
  >{\centering\arraybackslash}p{0.14\linewidth}
  >{\centering\arraybackslash}p{0.14\linewidth}
}
\hline
\ttcode{N\_between\_samples} & \textbf{1} & \textbf{10} & \textbf{100} & \textbf{500} & \textbf{1000} \\
\textbf{$\tau_\mathrm{int}$} (energy) & 1895 & 141.4 & 19.7 & 3.24 & 1.64 \\
\hline
\end{tabularx}
\end{center}

For very small \ttcode{N\_between\_samples}, $\tau_\mathrm{int}$ is very high: in cases where the update is rejected, the configuration is identical to the one measured before! A choice of \ttcode{N\_between\_samples} between 500 and 1000 would be a good tradeoff between $\tau_\mathrm{int}$ and runtime for this example. Increasing \ttcode{N\_between\_samples} to well over 1000 would be a waste of CPU time.

The second setup illustrates critical slowing down. We use the Python CLI and sweep across a continuous topological phase transition on the square lattice for $L \in \{ 8,12,16\}$ (see Sec.~\ref{sec:top_pt}):
\begin{lstlisting}[language=bash, basicstyle=\ttfamily\small\color{darkblue}, breaklines=true, caption={Critical slowing down benchmarking setup}]
python3 -u ./python/cli/paratoric.py -sim etc_h_sweep -Ns 100000 -Nbs 160*L^2 -Nth 10000*L^2 -T 1/L -muc 1 -hl 0.2 -hu 0.5 -hs 64 -hct 0.0 -Jc 1 -lmbdac 0.2 -lmbdact 0.2 -Nr 1000 -cth 0 -obs percolation_strength percolation_probability plaquette_percolation_strength plaquette_percolation_probability largest_cluster largest_plaquette_cluster string_number energy energy_h energy_mu energy_J energy_lmbda sigma_x sigma_z star_x plaquette_z staggered_imaginary_times delta anyon_count anyon_density fredenhagen_marcu sigma_x_static_susceptibility sigma_x_dynamical_susceptibility sigma_z_static_susceptibility sigma_z_dynamical_susceptibility -s 0 -bas x -lat square -L 16 -bound periodic -dsp 1 -proc 64 -snap 0 -fts 0 -outdir /scratch/s/Simon.Linsel/toric_code/out
\end{lstlisting}

\begin{figure}[t]
\centering
\includegraphics[width=0.95\textwidth]{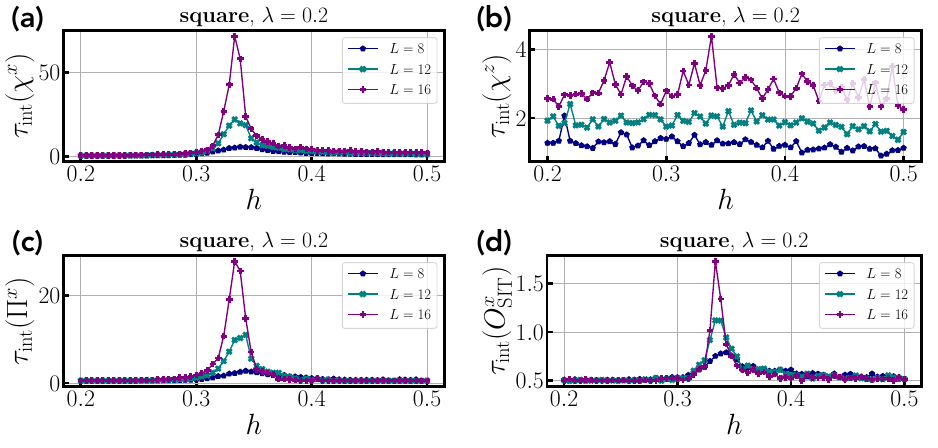}
\caption{\textbf{Critical slowing down} of the integrated autocorrelation times in the vicinity of a continuous (topological) phase transition. We simulate the extended toric code (\ref{eq:eTC}) on the square lattice in the $\hat{\sigma}^x$-basis around the critical field $h_\mathrm{c}(\lambda=0.2) \approx 0.33$ \cite{Wu2012, Linsel2024}. (a) The susceptibility $\chi^x$ features a maximum around the critical field. The growth of $\tau_\mathrm{int}$ with $L$ is superlinear: we observe critical slowing down. (b) The susceptibility $\chi^z$ does not have a maximum near the critical field and does not exhibit critical slowing down because the phase transition is driven by $h/J$ (not $\lambda/\mu$). (c) The percolation probability $\Pi^x$ is an order parameter and exhibits critical slowing down. (d) The staggered imaginary time order parameter $O^x_\mathrm{SI}$ also exhibits critical slowing down, albeit with \textit{much} smaller $\tau_\mathrm{int}$ than $\Pi^x$.}
\label{fig:Autocorrelation}
\end{figure}

In Fig.~\ref{fig:Autocorrelation}, we show $\tau_\mathrm{int}$ of $\chi^x$ (a), $\chi^z$ (b), $\Pi^x$ (c), and $O^x_\mathrm{SI}$ (d). All observables except $\chi^z$ can probe the phase transition. Unfortunately, they also exhibit critical slowing down, signaled by a maximum of $\tau_\mathrm{int}$ (around the critical field) which grows superlinearly with $L$. Our results nicely demonstrate how different observables can have vastly different $\tau_\mathrm{int}$. In particular, $O^x_\mathrm{SI}$ has a very low $\tau_\mathrm{int}$, which is very convenient for run-times.

$\chi^z$ cannot probe the phase transition as it is driven by $h/J$ (not $\lambda/\mu$). It does neither exhibit critical slowing down, nor even a maximum of $\tau_\mathrm{int}$ around the critical field. $\tau_\mathrm{int}$ has an approximately linear increase with $L$ for $\chi^z$ (and for $\chi^x$ far away from the phase transition) because in \ttcode{N\_between\_samples} we did not take into account the growth of the imaginary time dimension ($\beta=L$). The susceptibilities are integrals over the imaginary time, so the autocorrelation gets worse. A simple fix is to add an additional $\beta$-dependency to \ttcode{N\_between\_samples}.

\subsubsection{Run-time}

We benchmark the run-time for two realistic parameter sets on the square lattice and varying system size. The first setup simulates the toric code without fields:\footnote{All tests were run on a laptop; some conditions, like the CPU temperature, were not identical for all simulations. The benchmarks are therefore only an approximation.}
\begin{lstlisting}[language=python, basicstyle=\ttfamily\small\color{darkblue}, caption={$L$ benchmarking setup 1}]
>>> import numpy as np
>>> from paratoric import extended_toric_code as etc
>>> L=20
>>> series, mean, std, binder, binder_std, tau_int = etc.get_sample(N_samples=10000, N_thermalization=500*L*L*L, N_between_samples=8*L*L*L, N_resamples=1000, custom_therm=False, observables=["energy", "sigma_x", "sigma_z"], seed=0, mu=1.0, h=0.0, h_therm=0.0, J=1.0, lmbda=0, lmbda_therm=0.0, basis='x', lattice_type="square", system_size=L, beta=L, boundaries="periodic", default_spin=1, save_snapshots=False)
\end{lstlisting}

We only run one test per system size. The results are:
\begin{center}
\small
\begin{tabularx}{\linewidth}{
  >{\raggedright\arraybackslash}p{0.24\linewidth}
  >{\centering\arraybackslash}p{0.14\linewidth}
  >{\centering\arraybackslash}p{0.14\linewidth}
  >{\centering\arraybackslash}p{0.14\linewidth}
  >{\centering\arraybackslash}p{0.14\linewidth}
  >{\centering\arraybackslash}p{0.14\linewidth}
}
\hline
$L$ & \textbf{4} & \textbf{8} & \textbf{12} & \textbf{16} & \textbf{20} \\
Runtime (s) & 3.1 & 21.3 & 75 & 197 & 379 \\
\hline
\end{tabularx}
\end{center}
From our experience, for large $L$ the update complexity is approximately $\mathcal{O}(L^3 \log \beta)$, owing to the chosen cubic dependency of \ttcode{N\_thermalization} and \ttcode{N\_between\_samples} and a $\mathcal{O}(\log \beta)$ dependence of operations on the imaginary time axis, see $\beta$ benchmark below. The system size itself does not impact the performance, as the interactions are local. On computing clusters, we have realized system sizes of up to $L=80$ for the square lattice; this number will only increase in the future as CPUs get faster.

The second setup simulates the toric code with fields in both $\hat{\sigma}^x$ and $\hat{\sigma}^z$-direction:
\begin{lstlisting}[language=python, basicstyle=\ttfamily\small\color{darkblue}, caption={$L$ benchmarking setup 2}]
>>> import numpy as np
>>> from paratoric import extended_toric_code as etc
>>> L=20
>>> series, mean, std, binder, binder_std, tau_int = etc.get_sample(N_samples=10000, N_thermalization=500*L*L*L, N_between_samples=8*L*L*L, N_resamples=1000, custom_therm=False, observables=["energy", "sigma_x", "sigma_z"], seed=0, mu=1.0, h=0.2, J=1.0, lmbda=0.2, basis='x', lattice_type="square", system_size=L, beta=L, boundaries="periodic", default_spin=1, save_snapshots=False)
\end{lstlisting}

We only run one test per system size. The results are:

\begin{center}
\small
\begin{tabularx}{\linewidth}{
  >{\raggedright\arraybackslash}p{0.24\linewidth}
  >{\centering\arraybackslash}p{0.14\linewidth}
  >{\centering\arraybackslash}p{0.14\linewidth}
  >{\centering\arraybackslash}p{0.14\linewidth}
  >{\centering\arraybackslash}p{0.14\linewidth}
  >{\centering\arraybackslash}p{0.14\linewidth}
}
\hline
$L$ & \textbf{4} & \textbf{8} & \textbf{12} & \textbf{16} & \textbf{20} \\
Runtime (s) & 3.9 & 34.1 & 133 & 323 & 689 \\
\hline
\end{tabularx}
\end{center}

We also test the run-time dependence of the inverse temperature $\beta$, with the following setup:
\begin{lstlisting}[language=python, basicstyle=\ttfamily\small\color{darkblue}, caption={$\beta$ benchmarking setup}]
>>> import numpy as np
>>> from paratoric import extended_toric_code as etc
>>> series, mean, std, binder, binder_std, tau_int = etc.get_sample(N_samples=10000, N_thermalization=20000, N_between_samples=2000, N_resamples=1000, custom_therm=False, observables=["energy", "sigma_x", "sigma_z"], seed=0, mu=1.0, h=0.2, J=1.0, lmbda=0.2, basis='x', lattice_type="square", system_size=10, beta=20, boundaries="periodic", default_spin=1, save_snapshots=False)
\end{lstlisting}

We only run one test per $\beta$. The results are:
\begin{center}
\small
\begin{tabularx}{\linewidth}{
  >{\raggedright\arraybackslash}p{0.24\linewidth}
  >{\centering\arraybackslash}p{0.14\linewidth}
  >{\centering\arraybackslash}p{0.14\linewidth}
  >{\centering\arraybackslash}p{0.14\linewidth}
  >{\centering\arraybackslash}p{0.14\linewidth}
  >{\centering\arraybackslash}p{0.14\linewidth}
}
\hline
$\beta$ & \textbf{4} & \textbf{8} & \textbf{12} & \textbf{16} & \textbf{20} \\
Runtime (s) & 14.9 & 17.2 & 19.1 & 20.0 & 22.1 \\
\hline
\end{tabularx}
\end{center}

This benchmark illustrates an appealing feature of our implementation: there is almost no slowing down when increasing $\beta$ implying that very low temperatures are within reach with ParaToric. This paradoxical result (because the number $n$ of off-diagonal star/plaquette and magnetic field operators must physically scale linearly in $\beta$) is explained by the fact that most searches within the imaginary time axis scale as $\mathcal{O}(\log n)$ by making use of binary searches.

\subsubsection{Topological phase transition} \label{sec:top_pt}

\begin{figure}[t]
\centering
\includegraphics[width=0.95\textwidth]{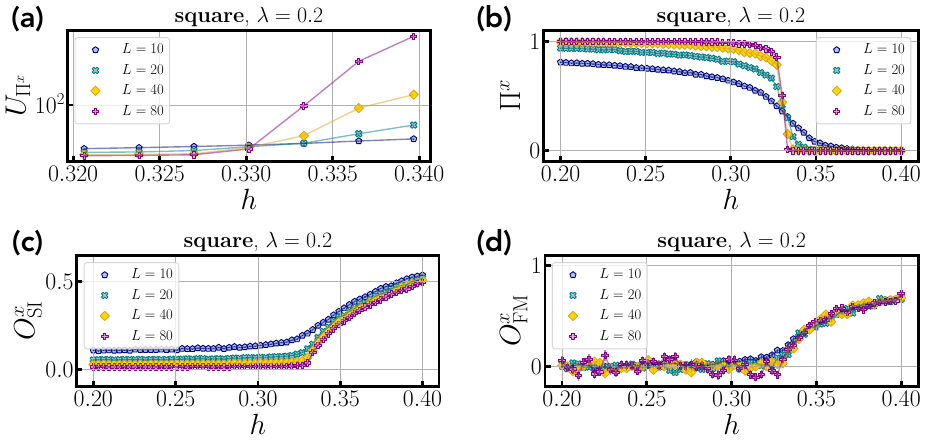}
\caption{\textbf{Topological phase transition} of the extended toric code (\ref{eq:eTC}) on the square lattice. The critical field is known and located at $h_\mathrm{c}(\lambda=0.2) \approx 0.33$ \cite{Wu2012, Linsel2024}. Our results agree with the value published in the literature, within error bars. (a) The percolation probability Binder ratio $U_{\Pi^x}$ \cite{Linsel2024} features a crossing point around $h=0.33$. (b) The percolation probability $\Pi^x$ is non-zero in the topological phase and zero in the trivial phase. The transition gets sharper with increasing system size. (c) The staggered imaginary time order parameter $O^x_\mathrm{SI}$ is zero in the topological phase and non-zero in the trivial phase. (d) The Fredenhagen-Marcu order parameter $O^x_\mathrm{FM}$ is zero in the topological phase and non-zero in the trivial phase. The loop length grows with $\mathcal{O}(L)$. Compared to the other order parameters, it is very noisy because it is a multi-body correlator and on top of that a division of two exponentially small numbers.}
\label{fig:TOPPT}
\end{figure}

We probe the well-known topological phase transition in the ground state of the extended toric code (\ref{eq:eTC}) on the square lattice, where we have a gapped $\mathbb{Z}_2$ quantum spin liquid for small fields $h,\lambda$ and a topologically trivial phase for high fields separated by a continuous phase transition. We set $J=\mu=1, \;\lambda = 0.2$ and sweep $h$ over the known critical value $h_\mathrm{c}(\lambda=0.2) \approx 0.33$ \cite{Vidal2009, Dusuel2011, Wu2012, Vanderstraeten2017, Crone2020, Linsel2024, Xu2024} for $L\in \{10,20,40,80\}$ in the $\hat{\sigma}^x$-basis. The temperature is set to $T=1/L$ to capture ground state physics. We take 30000 snapshots, with $8 L^3$ steps in between snapshots and $500 L^3$ thermalization steps.\footnote{If we were interested in quantities like critical exponents and we need to go very close to the critical field, we should take into account the dynamical exponent $z$ ($\tau_\mathrm{int} \sim L^z$) in the number of steps between snapshots to account for critical slowing down. Now the error bars are just larger near the critical field.}. We confirm that the systems are well thermalized. We show the percolation probability, the Fredenhagen-Marcu string order parameter, and the staggered imaginary time order parameter in Fig.~\ref{fig:TOPPT}. All of them reproduce the known phase boundary.

\section{Conclusion \& Outlook}
We have presented ParaToric, a continuous-time quantum Monte Carlo solver for the toric code in a parallel field. ParaToric builds on the existing work of Wu, Deng, and Prokof'ev \cite{Wu2012} and is also applicable to high temperature and low off-diagonal couplings.

ParaToric can store snapshots, which makes it ideally suited to generate training/benchmarking data for applications in other fields, such as lattice gauge theories, cold atom or other quantum simulators, quantum spin liquids, artificial intelligence, and quantum error correction. We believe it also serves a pedagogical purpose. Another strength of ParaToric is its interoperability with other programming languages. The C interface is compatible with virtually all programming languages, thus ParaToric can be seamlessly integrated into other projects.

ParaToric comes with an MIT license. For future releases of ParaToric we plan extensions along the following lines:
\begin{itemize}
    \item Additional lattices such as the kagome and the ruby lattice. Given the underlying graph structure used in ParaToric, such extensions are straightforward.
    \item Additional observables: we think here of, for instance, the finite temperature extension of the fidelity susceptibility to diagnose the phase transitions in the absence of a local order parameter. It would be worthwhile to have additional off-diagonal observables such as the off-diagonal Fredenhagen-Marcu string operators, or correlation functions between off-diagonal operators in space and or time. Measurements of the Rényi entropy are also high on the to-do list. The latter two classes require however major changes to the code, and testing.
    \item Additional interaction types. There are many classes of models in which topological order may be emergent instead of explicit as in the toric code. Such models typically have additional interactions than the ones covered in ParaToric, such as longer-range Ising interactions, and miss some others (typically the plaquette type interactions, and sometimes even the star terms). It is in general an open problem how to efficiently simulate such models at the lowest temperatures (even for sign-free models). Frustrated interactions (both in toric code variants as well as in dual models) are also an important class with many unsolved problems. Extending ParaToric to dealing with other types of interactions can thus serve as an additional tool for benchmarking purposes and algorithmic exploration.
\end{itemize}

\FloatBarrier

\section*{Acknowledgements}
The authors acknowledge fruitful discussions with A. Bohrdt, G. De Paciani, G. Dünnweber, F. Grusdt, L. Homeier, and N. V. Prokof'ev.

\paragraph{Author contributions}
SML did the main coding and planning work with input from LP. All authors contributed to the writing of the manuscript.

\paragraph{Funding information}
This research was funded by the European Research Council (ERC) under the European Union’s Horizon 2020 research and innovation program -- ERC Starting Grant SimUcQuam (Grant Agreement No. 948141), and by the Deutsche Forschungsgemeinschaft (DFG, German Research Foundation) under Germany's Excellence Strategy -- EXC-2111 -- project number 390814868. The project/research is part of the Munich Quantum Valley, which is supported by the Bavarian state government with funds from the Hightech Agenda Bayern Plus. L.P. acknowledges financial support from ANR-23-CE30-0018.




\end{document}